%% file: INetCEP-main-IEEE.tex
\documentclass[journal]{IEEEtran}
\usepackage{comment}
\usepackage[utf8x]{inputenc}
\usepackage{amsmath,amssymb,amsfonts}
\usepackage{algorithmic}
\usepackage{graphicx}
\usepackage{textcomp}
\usepackage{xcolor}
\usepackage{todonotes}
\usepackage{paralist}
\usepackage{numprint}
\usepackage{xspace}
\usepackage{hyperref} 
\usepackage{cleveref}
\usepackage{multirow}
\usepackage{subcaption}
\usepackage{soul}
\usepackage{balance}
\usepackage[acronym,toc,shortcuts]{glossaries}
\usepackage{listings}
\usepackage{float}
\PassOptionsToPackage{normalem}{ulem}
\usepackage{ulem}
\usepackage{bytefield} 
\usepackage{booktabs}
\providecolor{added}{rgb}{0,0,1}
\providecolor{deleted}{rgb}{1,0,0}

\restylefloat{table}
\usepackage{fancyhdr} 
\usepackage[linesnumbered,ruled,vlined,lined,boxed]{algorithm2e}
\usepackage{euscript} 

\usepackage{scalerel}
\usepackage{tikz}
\usetikzlibrary{svg.path}

\definecolor{orcidlogocol}{HTML}{A6CE39}
\tikzset{
  orcidlogo/.pic={
    \fill[orcidlogocol] svg{M256,128c0,70.7-57.3,128-128,128C57.3,256,0,198.7,0,128C0,57.3,57.3,0,128,0C198.7,0,256,57.3,256,128z};
    \fill[white] svg{M86.3,186.2H70.9V79.1h15.4v48.4V186.2z}
                 svg{M108.9,79.1h41.6c39.6,0,57,28.3,57,53.6c0,27.5-21.5,53.6-56.8,53.6h-41.8V79.1z M124.3,172.4h24.5c34.9,0,42.9-26.5,42.9-39.7c0-21.5-13.7-39.7-43.7-39.7h-23.7V172.4z}
                 svg{M88.7,56.8c0,5.5-4.5,10.1-10.1,10.1c-5.6,0-10.1-4.6-10.1-10.1c0-5.6,4.5-10.1,10.1-10.1C84.2,46.7,88.7,51.3,88.7,56.8z};
  }
}

\newcommand\orcidicon[1]{\href{https://orcid.org/#1}{\mbox{\scalerel*{
\begin{tikzpicture}[yscale=-1,transform shape]
\pic{orcidlogo};
\end{tikzpicture}
}{|}}}}

\usepackage{amsthm} 
\theoremstyle{definition}

\theoremstyle{remark}

\newcommand\footnoteref[1]{\protected@xdef\@thefnmark{\ref{#1}}\@footnotemark}
\setdefaultenum{(1)}{}{}{}
\SetKwProg{Function}{function}{}{end function}
\SetKwProg{Upon}{upon}{ do}{trigger}
\SetKwInOut{KwVar}{Variables}
\SetKwInOut{KwDescribe}{Description}
\SetKw{KwTrigger}{trigger}
\SetKw{KwTo}{to}
\SetKw{KwAnd}{and}
\SetKwIF{If}{ElseIf}{Else}{if}{then}{else if}{else}{end if}
\SetKwFor{For}{for}{do}{end for}
\SetKwFor{ForEach}{for each}{do}{end for}
\SetKwFor{ForAll}{for all}{do}{end for}
\SetKwFor{ForAllP}{for all}{do in parallel}{end for}

\def\BibTeX{{\rm B\kern-.05em{\sc i\kern-.025em b}\kern-.08em
    T\kern-.1667em\lower.7ex\hbox{E}\kern-.125emX}}

\newcommand{\mi}[1]{\mathit{#1}}

\newcommand{\system}{\textsc{INetCEP}\xspace}
\newcommand{\Data}{\texttt{Data}\xspace}
\newcommand{\Interest}{\texttt{Interest}\xspace}
\newcommand{\DataStream}{\texttt{Data Stream}\xspace}
\newcommand{\QInterest}{\texttt{Continuous Interest}\xspace}
\newcommand{\AddQInterest}{\texttt{Add Continuous Interest}\xspace}
\newcommand{\RmQInterest}{\texttt{Remove Continuous Interest}\xspace}

\newcommand{\eg}{e.g., }
\let\OldS\S
\renewcommand{\S}{\OldS\xspace}
\Crefformat{figure}{#2Fig.~#1#3}
\Crefmultiformat{figure}{Figs.~#2#1#3}{ and~#2#1#3}{, #2#1#3}{ and~#2#1#3}

\newcommand*\mean[1]{\bar{#1}} 

\makeglossaries
\newacronym{CEP}{CEP}{Complex Event Processing}
\newacronym{NFN}{NFN}{Named Function Networking}
\newacronym{IoT}{IoT}{Internet-of-Things}
\newacronym{NDN}{NDN}{Named Data Networking}
\newacronym{ICN}{ICN}{Information-centric Networking}
\newacronym{CCN}{CCN}{Content-centric Networking}
\newacronym{PIT}{PIT}{Pending Interest Table}
\newacronym{FIB}{FIB}{Forwarding Information Base}
\newacronym{CS}{CS}{Content Store}
\newacronym{OP}{OP}{Operator Placement}
\newacronym{QoS}{QoS}{Quality of Service}
\newacronym{INP}{INP}{in-network processing}

\newboolean{komversion}
\setboolean{komversion}{true} 
\ifthenelse{\boolean{komversion}}{
    \usepackage{eso-pic}
    \AddToShipoutPictureBG{
        \AtPageUpperLeft{%
            \raisebox{-3\baselineskip}{
\makebox[\paperwidth]{\colorbox{yellow!40}{\begin{minipage}{0.85\paperwidth}\centering\small
                            {Manisha Luthra et al., \emph{Efficient Complex Event Processing in Information-centric Networking at the Edge.} This is a preprint of the paper submitted at IEEE IoT Journal.}
                        \end{minipage}}}}%
                    }
                    \AtPageLowerLeft{%
                        \raisebox{1.7\baselineskip}{
\makebox[\paperwidth]{\fbox{\begin{minipage}{0.92\paperwidth}\scriptsize
                                        {The documents distributed by this server have been provided by the contributing authors as a means to ensure timely dissemination of scholarly and technical work on a non-commercial basis. Copyright and all rights therein are maintained by the authors or by other copyright holders, not withstanding that they have offered their works here electronically. It is understood that all persons copying this information will adhere to the terms and constraints invoked by each author's copyright. These works may not be reposted without the explicit permission of the copyright holder.}
                                    \end{minipage}}}}%
                    }
    }
}{}
\begin{document}

\title{Efficient Complex Event Processing in Information-centric Networking at the Edge
}

\author{Manisha Luthra\orcidicon{0000-0002-3788-6664}\, \IEEEmembership{Student Member, IEEE},
Johannes Pfannmüller,
Boris Koldehofe\orcidicon{0000-0002-1588-2056}\,
\IEEEmembership{Member, IEEE}
Jonas Höchst\orcidicon{0000-0002-7326-2250}
Artur Sterz\orcidicon{0000-0001-9820-7373}
Rhaban Hark\orcidicon{0000-0003-1799-0023}
\IEEEmembership{Member, IEEE}
and
Bernd Freisleben\orcidicon{0000-0002-7205-8389}
\IEEEmembership{Member, IEEE}
\thanks{M. Luthra, B. Koldehofe, J. Höchst, A. Sterz, R. Hark and B. Freisleben are with the Department of Electrical and Computer Engineering at Technical University of Darmstadt, Germany (email: manisha.luthra@kom.tu-darmstadt.de).
J. Pfannmüller was with Technical University of Darmstadt and is with Software AG, Germany.
}
\thanks{Preprint submitted at IEEE IoT Journal on November 16, 2020.}
}

\maketitle

\begin{abstract}
\ac{ICN} is an emerging Internet architecture that offers promising features, such as in-network caching and named data addressing, to support the edge computing paradigm, in particular Internet-of-Things (IoT) applications.
ICN can benefit from Complex Event Processing (CEP), which is an in-network processing paradigm
to specify and perform efficient query operations on data streams. 
However, integrating CEP into ICN is a challenging task due to the following reasons:
(1) typical ICN architectures do not provide support for forwarding and processing continuous data streams;
(2) IoT applications often need short response times and require robust event detection, which both are hard to accomplish using existing CEP systems.

In this article, we present a novel network architecture, called \system, for
efficient CEP-based in-network processing
as part of \ac{ICN}.
\system enables efficient data processing in ICN
by means of
(1) a unified communication model that supports continuous data streams,
(2) a meta query language for CEP to specify data processing operations in the data plane, and
(3) query processing algorithms to resolve the specified operations.
Our experimental results for two IoT use cases and datasets show that \system offers very short response times of up to $73 \mu s$ under high workload and is more than $15 \times$ faster in terms of forwarding events than the state-of-the-art CEP system Flink.
Furthermore, the delivery and processing of complex queries is around $32\times$ faster than Flink and more than $100\times$ faster than a naive pull-based reference approach, while maintaining $100\%$ accuracy.
\end{abstract}


\input{sections/introduction}
\input{sections/background}

\input{sections/problem_wochanges}
\input{sections/design_wochanges}

\input{sections/evaluation}
\input{sections/discussion}

\balance
\bibliographystyle{abbrv}
\bibliography{INetCEP-main-IEEE}

\end{document}

%% file: sections/introduction.tex
\section{Introduction} \label{sec:intro}

\IEEEPARstart{I}{nformation-centric} Networking (ICN) is aimed at simplifying the principles of the Internet by changing its addressing scheme from \emph{named hosts} to \emph{named data}.
It focuses on the question \emph{what data is required} rather than \emph{where to get the data from}.
\ac{ICN} evolved into a key architecture for the edge computing paradigm towards realizing a content-centric Internet of the future~\cite{Ullah2020}.
Nowadays, \ac{ICN} is broadly adopted by academia and industry, \eg by Internet2, Cisco,
and Intel,
in real-world deployments~\cite{ICNDeploy2019}.
Since ICN inherently focuses on data, it particularly supports the utilization of in-network processing functionality, \eg to specify higher order functions on the data already within the network, as provided by \ac{CEP}.
\ac{CEP} is a powerful \emph{in-network processing} paradigm that takes a \emph{query} as input to describe correlations over incoming data streams in order to deliver \emph{complex events} in the form of data notifications in response to the \emph{query}~\cite{Buchmann2009}.
An example for a CEP application is a disaster scenario, where emergency management operations require a heat map of victim locations to better coordinate the activities of rescue workers~\cite{Alvarez2018}.
The heatmap can be produced efficiently using a CEP query in the network based on a set of incoming sensor data streams.
However, existing \ac{ICN} architectures like \ac{NDN}~\cite{Jacobson2009} and \ac{NFN}~\cite{Tschudin2013} lack mechanisms for in-network processing in the data plane. Particularly, these architectures are restricted in terms of (i) the underlying communication strategy, and (ii) higher level specification and data processing for \ac{IoT} applications.
Thus, \ac{ICN} architectures can benefit from the CEP paradigm, but deploying \ac{CEP} as part of an \ac{ICN} architecture is quite challenging.

A major challenge is the lack of support to forward and process incoming periodic data streams due to \ac{ICN}´s underlying communication strategy.
State-of-the-art \ac{ICN} architectures either rely on a \emph{pull}-based (\emph{consumer-initiated}) interaction pattern~\cite{Jacobson2009, Tschudin2014}
or a \emph{push}-based (\emph{producer-initiated}) interaction pattern~\cite{Chen11, Koponen2007}.
For periodic data streams, \emph{pull-based} interaction poses several problems: (1) it causes significant overhead in terms of request messages required to fetch each data object;
(2) it causes high delays until fresh data becomes available;
(3) polling for data might result into stale data served from an in-network cache;
(4) once a producer does not have any data to deliver, it is possible that the consumer ends up in repeatedly requesting data and busy-waiting for data to arrive, thus, unnecessarily allocating resources.

In contrast, changing to a \emph{push-based} interaction~\cite{Chen11} as supported by \ac{CEP} systems is problematic for  \ac{IoT} applications that are based on request-reply interactions. For example, traditional web applications still need personalized request-reply interaction.
Work on mitigating this problem~\cite{Ahmed2016} relies on  push-based beacon messages to inform consumers about arriving new data such that consumers can initiate a request-reply interaction.
Other researchers~\cite{Guendogan2018} integrate the push-based semantics in a centralized control plane.
However, such mechanisms -- besides adding complexity -- either incur a significant message overhead or might create a bottleneck for large-scale IoT applications.

Ideally, \ac{ICN} architectures should provide efficient support for both interaction patterns as part of a unified communication enabling any type of application.
Initial work in the context of content routing has shown the potential of a unified communication model~\cite{Carzaniga2011, Carzaniga2014}. Another proposal solves this problem by using long lived request packets~\cite{Guendogan2018}.
Although this approach provides scalable routing in a unified manner, the authors do not deal with the side-effect of enabling \emph{push-based} interaction in \ac{ICN}, i.e., \emph{flow imbalance}~\cite{ICNRG2019} resulting in event loss.
In the latter approach, long lived requests result in a large in-network state and might still lead to stale data.

We address these limitations by presenting a novel network architecture, called \system\footnote{This article is an extension of the authors' previous work~\cite{Luthra2019,INetCEPGithub2019}.}, for efficient CEP-based \emph{in-network processing} in combination with \emph{a unified communication model} within \ac{ICN}.
\system  enables \emph{efficient and in order} processing of data in the forwarding plane 
to support the requirements of IoT applications, such as
detecting complex events with short response times (as low as a few milliseconds) in a highly accurate manner, while potentially dealing with large numbers of events.

Our contributions are as follows:
\begin{enumerate}
\item We propose a unified communication model that provides coexisting pull- and push-based communication patterns to realize \ac{CEP} applications while still supporting classical request-reply interactions.
Additionally, we provide a rate-based flow control mechanism to mitigate the issue of flooded links to ensure \emph{flow balance}.
\item We introduce a meta query language to express complex event derivations on the data plane over the \ac{ICN} substrate.
\item We present centralized and distributed query processing algorithms that ensure efficient and in order processing of
queries in a distributed manner.
\item Finally, we present an open source implementation~\cite{INetCEPGithub2019} and evaluation of \system on state-of-the-art \ac{ICN} architectures, namely \ac{NFN}~\cite{Tschudin2014} and CCN-lite~\cite{CCN-lite2019}, using two \ac{IoT} case studies and open datasets from the DEBS Grand Challenge 2014~\cite{DEBS2014} and a disaster communication scenario~\cite{Alvarez2018}.
\end{enumerate}

Our evaluation results show that \system is more than \textbf{$15\times$} faster in terms of forwarding events compared to the state-of-the-art CEP system \emph{Flink}~\cite{Carbone2015ApacheFS}.
Furthermore, the delivery and processing of complex queries on the mentioned datasets is around \textbf{$32\times$} faster than \emph{Flink} and more than $100\times$ faster than a naive \emph{pull-based} reference approach while maintaining $100\%$ accuracy.

The paper is organized as follows.
In Section \ref{sec:preliminaries}, we present related work.
In Section \ref{sec:problem}, we describe the research challenges.
In Section \ref{sec:design}, we present the design of \system. In Section \ref{sec:eval}, we provide a comprehensive evaluation of \system.
Section \ref{sec:conclusion} concludes the paper and outlines areas for future work.

%% file: sections/background.tex
\section{Related Work}~\label{sec:preliminaries}

We provide background information and related work on the building blocks of our proposal: \ac{CCN} and Complex Event Processing (CEP) in the following.

\subsection{Content-centric Networking}~\label{subsec:CCN}
Jacobson et al.~\cite{Jacobson2009} proposed CCN\footnote{In the remainder of the paper, we will use the terms ICN and
CCN interchangeably.}, where communication is \emph{consumer-initiated}, consisting of two packets: \Interest  and \Data.
A data object (payload of a \Data packet) satisfies an interest if the name in the \Interest packet is a prefix of the name in the \Data packet.
Thus, when a packet arrives on a face\footnote{face stands for interface in \ac{CCN} terminology.} (identified by $face\_id$) of a CCN node, the longest prefix match is performed on the name and the data is returned based on a lookup.

\textit{CCN data plane:} Each \ac{CCN} node maintains three major data structures: \ac{FIB}, \ac{CS} (also known as in-network cache), and \ac{PIT}. Once an \Interest arrives on a face, the node first checks its Content Store for a matching \Data packet by name. Upon a match, the \Data packet is sent via the same face it arrived from. Otherwise, the node continues its search in the \ac{PIT} that stores all the \Interest packets (along with its incoming and outgoing face) that are not satisfied. If an entry exists in the \ac{PIT}, the face is updated and the \Interest is discarded, because an \Interest packet has already been sent upstream. Otherwise, the node looks for a matching \ac{FIB} entry and forwards the \Interest to the potential source(s) of the data. 

\textit{NDN and NFN:} \ac{NDN}~\cite{Zhang2014a} emerged as an architecture that is based on the principles of CCN's \emph{named data}.
\ac{NFN} is another emerging architecture that focuses on addressing \emph{named functions} in addition to \emph{named data} by extending the principles of \ac{NDN}.
NFN blends~\cite{Tschudin2014} data computations with network forwarding, by performing computational tasks across the CCN network. It represents \emph{named functions} on the data as
$\lambda-$calculus expressions.
We aim to encapsulate CEP operators (cf. next section) as NFN named functions and hence resolve them in the network.
However, NFN focuses mainly on resolving discrete functions on top of the CCN substrate. In contrast, we focus on continuous and discrete computations (push and pull), expressive representation of the computation tasks, and their efficient distribution (cf. Section \ref{sec:design}).

\subsection{Complex Event Processing}~\label{subsec:CEP}
CEP can process multiple online and unbounded data streams using compute units called \emph{operators} to deliver meaningful events to consumers~\cite{Buchmann2009}. The consumers specify interest in the form of a \emph{query} consisting of multiple operators.
 Some of the commonly used operators are defined below~\cite{Luthra2019}:
\begin{enumerate}
\item \texttt{\textbf{Filter}} checks a condition on the attribute of an event tuple and forwards the event if the condition is satisfied.
\item \texttt{\textbf{Aggregate}} applies an aggregation function such as $max$, $min$, $count$, $sum$, $avg$, etc., on one or more event tuples. Hence, the data stream must be bounded to apply these operations. For this purpose, \emph{window} can be used.
\item \texttt{\textbf{Window}} limits the unbounded data stream to a window based on time or tuple size, such that operators like \texttt{Aggregate} can be applied on the selected set of tuples\footnote{In this article, we consider sliding windows that are most widely used, although other windows, such as tumbling windows, have been proposed.}.
\item \texttt{\textbf{Join}} combines two data streams to one output stream based on a filter condition applied on a window of limited tuples.
\end{enumerate}

These operators can be \emph{stateless} or \emph{stateful}. \texttt{{Filter}} and \texttt{{Aggregate}} are stateless operators, while the other operators are stateful and maintain the state of input tuples before emitting the complex event and therefore depend on multiple input tuples to be accumulated before actual emission.

\input{sections/related-work}

%% file: sections/related-work.tex
\subsection{ICN Architectures}~\label{subsec:icnarch}
We classify ICN architectures based on their communication capabilities as pull-based, push-based, or both push- and pull-based.

\textbf{Pull-based.}
Classical architectures in ICN like \ac{NDN}~\cite{Zhang2014a} and \ac{NFN}~\cite{Tschudin2014} use a consumer-initiated communication strategy.
This essentially means that the consumer \emph{pulls} the data from the producer by sending an \Interest(request) packet to the network for each data item produced.
The \ac{NDN} network forwards this request to one or more producers that satisfy the request and then forward the \Data (reply) back to the consumer. Such architectures do not provide support for forwarding and processing of periodic data streams.
Several solutions building on top of NDN~\cite{Sifalakis2014, Krol2017} and NFN~\cite{Shang2016, Scherb2017, Tschudin2013} try to solve this problem by continuously polling the producer with \Interest packets. This approach works well when the producer sends the data stream at lower rates, but at higher event rates a lot of traffic is generated.
Furthermore, the consumer can miss data objects or receive stale data.

\textbf{Push-based.}
The second class of \ac{ICN} architectures~\cite{Koponen2007, Chen11} is based on  \emph{push-based} or producer-initiated communication similar to publish-subscribe systems. Such architectures lack support for request-reply interactions such as web applications. To solve this issue, some authors~\cite{Ahmed2016} use so-called beacon messages to support request-reply interactions. Another work solves this problem by using long lived request packets~\cite{Guendogan2018}. However, the former approach incurs a significant message overhead, while the latter approach poses a bottleneck for large-scale IoT applications.

\textbf{Both push and pull.}
Some approaches like CONVERGENCE~\cite{Melazzi2010}, GreenICN~\cite{GreenICN}, Carzaniga et al.~\cite{Carzaniga2011} and HoPP~\cite{Guendogan2018} provide support for both \emph{push-} and \emph{pull-} based communication strategies. The CONVERGENCE system combines the publish/subscribe interaction paradigm on top of an information-centric network layer. In contrast, we provide a unified interface such that pull and push based interaction patterns can co-exist in a single network layer while performing in-network computations.
GreenICN is an \ac{ICN} architecture by combining NDN (pull-based) with COPSS~\cite{Chen11} (push-based). However, their focus is only on post-disaster~\cite{Tagami2016} and video streaming scenarios~\cite{GreenICN}, while we focus on a wider range of IoT applications. Due to the limited range of scenarios, their focus also is not on improving response times. 
Carzaniga et al.~\cite{Carzaniga2011} present a unified network interface similar to our work, but the authors only propose a preliminary design of their approach without implementing it in an \ac{ICN} architecture, and subsequently focus on routing decisions~\cite{Carzaniga2013} rather than on distributed processing.
HoPP~\cite{Guendogan2018} is closest to our approach in implementing coexisting push and pull functionality, but the authors have moved the push semantics to a centralized control plane which could be a bottleneck for a large scale IoT network.

\subsection{\ac{CEP} and Networking Architectures}~\label{subsec:eparch}
We now discuss relevant work on CEP, emerging networking architectures, and data plane languages.

\textbf{\ac{CEP} Architectures.} Modern event processing architectures support low latency, e.g., Apache Flink~\cite{Carbone2015ApacheFS}, Twitter's Heron~\cite{Kulkarni2015}, and Google's Millwheel~\cite{Akidau2013}. However, these architectures operate in a middleware layer while we propose to use the same functionality in the data plane with ultra-low latency in terms of forwarding.
Another possibility is to interface event processing architectures with an \ac{ICN} architecture. Initial work implemented Hadoop on \ac{NDN}~\cite{Gibbens2017} for datacenter applications. However, this requires changing the network model to push, which would limit the support for request-reply interactions.
Reliable CEP architectures~\cite{Koldehofe2013, Ottenwalder2014} offer reliable event processing with minimal overhead, for instance by checkpointing~\cite{Koldehofe2013} and using buffers~\cite{Ottenwalder2014}. However, they suffer from high latencies either due to the round trip to the overlay network or by the inherent overhead introduced by the reliability mechanisms.

\textbf{Networking Architectures.} Another network architecture is Software-Defined Networking (SDN)~\cite{Kreutz2015}, which is gradually being deployed, \eg in Google's data centers. It allows network managers to program the control plane to support efficient traffic monitoring and engineering. The {SDN} architecture is complementary to our work, since SDN empowers the control plane, while \ac{ICN} upgrades the data plane of the current Internet architecture.

\textbf{Data Plane and Query Languages.}
Many CEP systems propose a query language that supports users to specify and embed the features provided in their applications~\cite{Weisenburger2017}. A particular, challenge when supporting network architectures is to be able to decompose the query specification such that it can be mapped to the programming models provided by dedicated network architectures.
The novelty of our proposed query language is to provide a mapping of operations to \ac{ICN}'s data plane.
Alternative designs build on P4
~\cite{Bosshart2014}.
Initial work on programming \ac{ICN} with P4 faced several difficulties due to lack of language features and the strong coupling of the language to SDN's data plane model~\cite{Signorello2016}.

%% file: sections/problem_wochanges.tex
\section{Research Challenges}\label{sec:problem}

We have three main objectives: (1) with coexisting \emph{pull-} and \emph{push-} based communication strategies, we aim to enable wide range of IoT applications over ICN architectures; (2) we aim to provide low response times for IoT applications, while maintaining throughput; (3) finally, we aim to provide support for lossless communication, robust event processing, and hence highly accurate complex events.

To better understand these objectives, consider as an example smart plugs that are currently used with devices like Amazon Alexa and Google Assistant, expected to be a \$62 million market by 2023~\cite{PMR2020}.
One of the important features of smart plugs is the information they provide to derive powerful load predictions for power grid providers.
By constantly predicting power usage, providers can put contingency measures in place to manage huge changes in demand in real time without any interruptions. These operations are typically managed by control networks, where a machine connected to a control network performs operations in place of a human. Such operations are
time-sensitive and need to be highly robust. In context of  load prediction, the machine might have to feed more power into the grid to prevent it from collapsing.
Such scenarios need continuous push notifications of load predictions and, most importantly, the control networks require short response times to plan for an action.
Additionally, smart plugs must be able to take requests from the end users. such as voice instructions by means of an app in conjunction with Alexa or Google assistant.
The example shows that \ac{CCN} has to support both types of communication strategies, pull- and push-bashed, at the same time.

Beyond coexisting communication strategies, the \ac{CCN} architecture must provide means to consumers to specify and represent \emph{queries} to correlate data from multiple primitive event producers and  process the data.
For example, to specify load predictions on the events generated by smart plugs, there must be a way to specify those predictions in a data plane query language and means to derive them.

The third problem of \ac{IoT} applications on top of \ac{CCN} is the requirement of highly \emph{robust} and \emph{scalable executions} while providing short response time.
As an example, smart plugs generate millions or even billions of power readings that have to be processed simultaneously.
There is a need to process this scale of events efficiently in the data plane without overloading the core network operations of a \ac{CCN}.
Furthermore, event loss is not acceptable in control networks. Even a single wrong instruction may lead to power outage in the smart plug example.

To achieve our objectives, we address two core challenges:

\textbf{R1: Continuous data streams.}
Currently, there is no \ac{ICN} architecture that supports coexisting push- and pull-based communication patterns in a single architecture.
The fundamental reason why \ac{ICN} architectures such as \ac{CCN} are \emph{pure-pull}-based
is their implicit flow control.
The one-to-one mapping of \Interest and \Data packets along with the PIT table ensures flow balance in an \ac{ICN} network~\cite{ICNRG2019}.
By introducing continuous data streams, the flow imbalance between the producer and the broker network could cause congestion and data loss.
Although simple flow control mechanisms, \eg restricting the sending rate, are valid countermeasures, they cannot detect and actively react to packet loss,
which is crucial for many IoT applications.

\textbf{R2: Efficient and in-order stream processing.}
The push nature of continuous data streams introduces the challenge of providing  efficient and in-order processing of data in the forwarding plane of \ac{CCN}.
First, the outcome of the CEP operators depends on the original order of events in the data stream as they are produced.
Hence, efficient and parallel processing of operators requires coordination among the processing threads. 
Second, large numbers of events and queries require the distribution of the processing queries into graphs of operators, which leads to the well-known operator placement problem~\cite{Nardelli2019, Luthra2018}.
The deployment of operators in a \ac{CCN} network introduces new challenges in terms of fulfillment of Quality-of-Service (QoS) metrics like short response times, throughput, and correctness in delivery of events.

%% file: sections/design_wochanges.tex
\begin{figure*}
	\centering
	\includegraphics[width=0.4\linewidth]{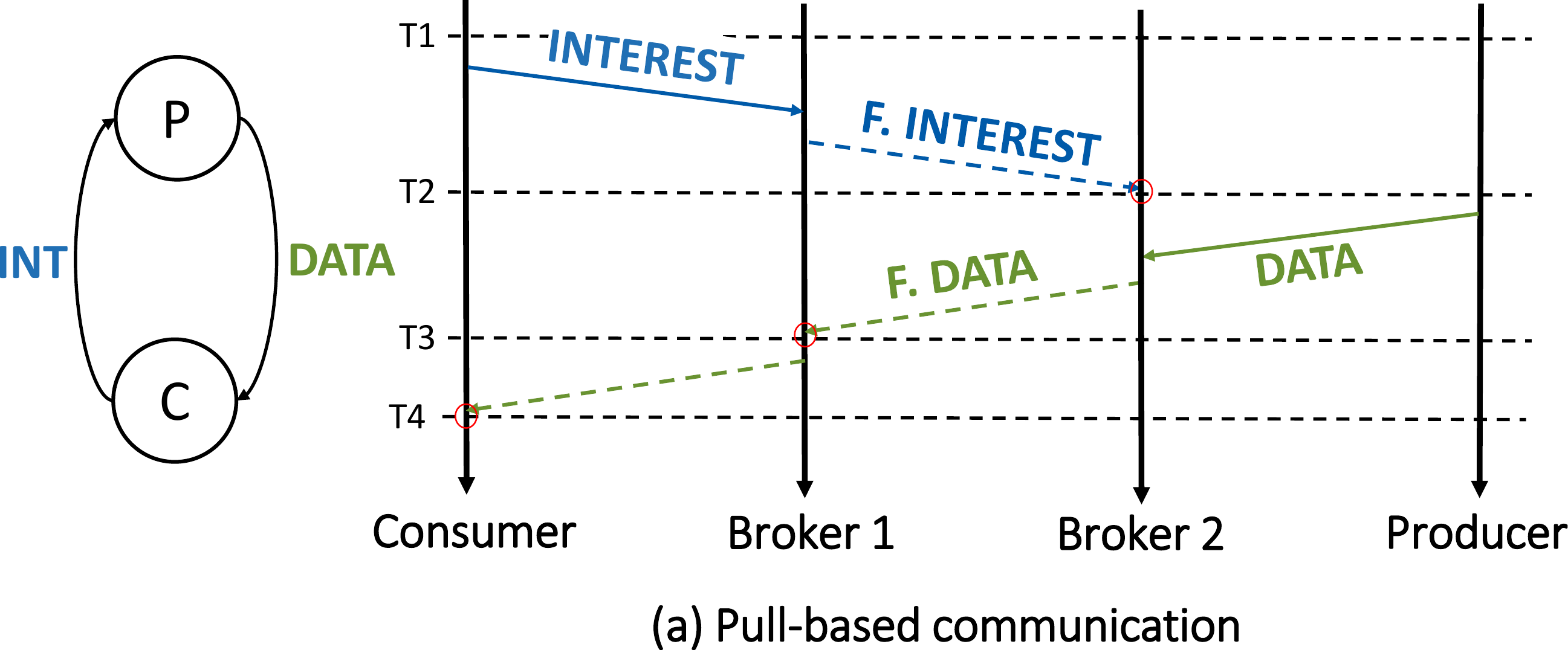}
	\includegraphics[width=0.38\linewidth]{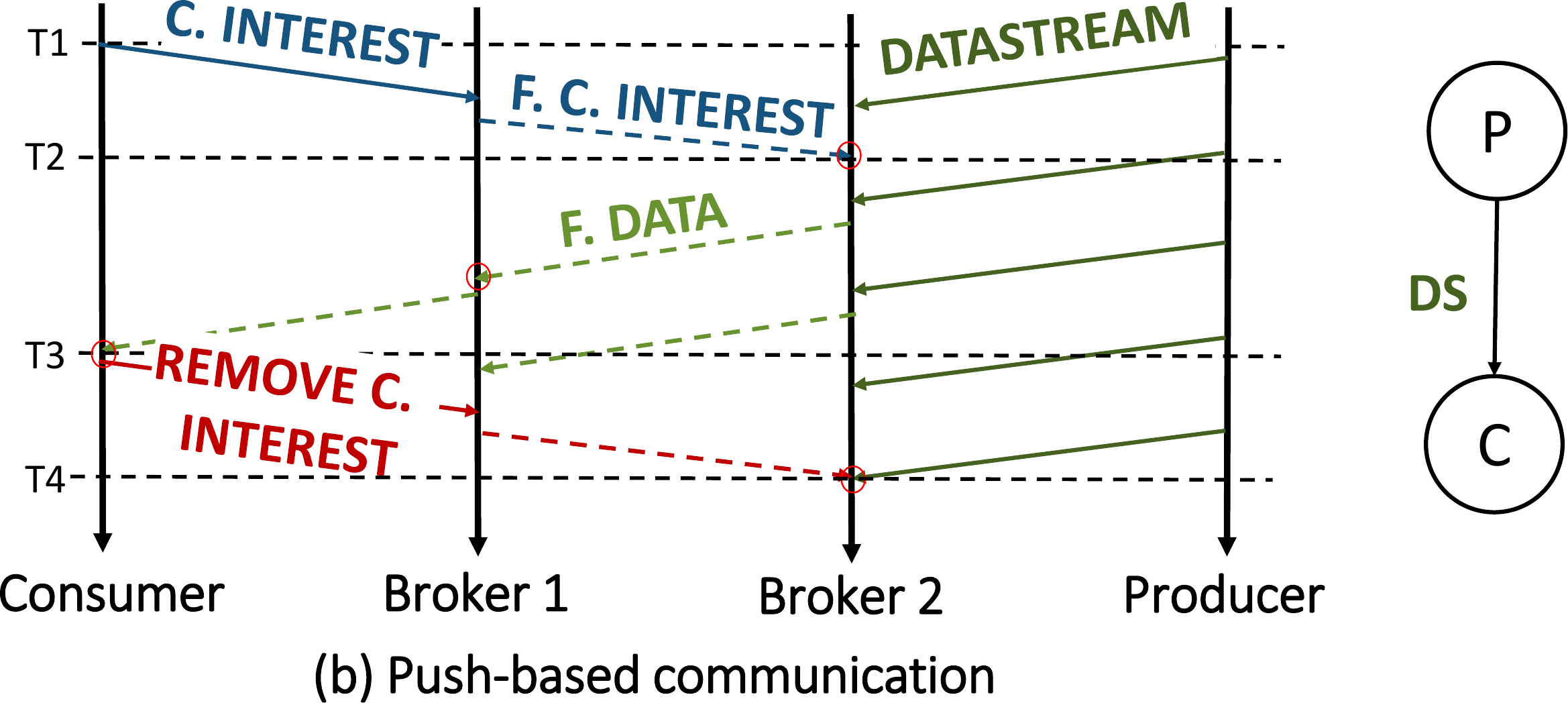}
	\setlength{\abovecaptionskip}{5pt}
	\setlength{\belowcaptionskip}{-17pt}
	\caption{\system communication model supports both push- and pull-based communication while maintaining a single PIT entry and fetching latest data object for the query.}
	\label{fig:inetcep}
\end{figure*}
\section{The \system Architecture}\label{sec:design}

In this section, we present the design of \system. We first discuss the core ideas of \system and then provide details of its architectural components.

\subsection{Core Ideas} \label{sec:ideas}
The \system architecture address the challenges discussed in the previous section based on the following two core ideas:

\subsubsection{A unified communication layer}
We complement the current \ac{CCN} architecture with three additional packet types to support the handling of periodic data streams.
We introduce the forwarding plane logic for the newly proposed types and propose algorithms for their handling.
To enable flow control, we use a lossless rate-based flow control mechanism~\cite{NAP1997} that balances the rate by using active feedback for the producer to adjust the sending rate based on the available resources.
This ensures no reduction in response time and prevents packet loss.
\par

\subsubsection{Reactive and asynchronous handling of data streams}
To provide efficient and in-order processing of periodic data streams in the forwarding plane, we propose algorithms for centralized and distributed query processing.
We perform the decomposition of queries into primitive operations that are executed in the CCN forwarding plane.
An operator placement service coordinates the deployment of operators onto paths that ensure the maximal rate and minimal response time in the delivery of events.
The processing of the operators is planned by asynchronous parallel threads
while maintaining so-called sub interests in the PIT of the placed nodes.
We offer coordination of threads to guarantee in-order processing.
The lossless rate based flow control mechanism along with in-network caching mechanism of CCN ensures correctness in the detection of complex events.

\input{sections/systemmodel_wo}
\begin{table*}
\scriptsize
\begin{tabular}{|p{2cm}|lp{4cm}|p{2cm}p{6cm}|}
\hline
\multirow{2}{*}{\textbf{Characteristic}}            & \multicolumn{2}{c|}{\textbf{Old Architecture}}                                       & \multicolumn{2}{c|}{\textbf{Our Architecture}}                                                                     \\
                                 & \ac{ICN} & Description                                     & \system           & Description                                                            \\ \hline
\multirow{5}{*}{\begin{minipage}{1.5cm}Packet Types (cf. \S\ref{subsubsec:handling})\end{minipage}}      & \Interest  & {Interest packet to express a request on a data object}                                & \Interest         & {Interest packet to express a request on a data object}    \\
                                 & \Data      &    {Data packet to satisfy a request}                              & \Data             & {Data packet to satisfy a request by an \Interest or \AddQInterest}                                                       \\
                                 & -                         & -                                               & \DataStream       & {Data Stream packet representing continuous time series data}                          \\
                                 & -                         & -                                               & \AddQInterest     & {Continuous Interest packet to express a continuous interest in the data object}                          \\
                                 & -                         & -                                               & \RmQInterest      & {Remove Continuous Interest packet to remove interest on a data object}                         \\ \hline
\multirow{3}{*}{\begin{minipage}{2cm}Data Structures \\ (cf. \S\ref{subsubsec:components})\end{minipage}} & \ac{PIT} & Stores pending interests of consumer            & \ac{PIT}        & Stores pending interests and continuous interests                           \\
                                 & \ac{CS}  & Stores data objects                             & \ac{CS}         & Stores data objects and buffers the data stream for stateful operators like window. \\
                                 & \ac{FIB} & Stores forwarding information towards producers & \ac{FIB}        & Stores forwarding information towards producers and consumers interested in CEP query                      \\ \hline
{\begin{minipage}{2cm}Data Processing  \\(cf. \S\ref{subsec:QE})  \end{minipage}}                & -                         & -                                               & \ac{CEP} engine & Parse, place, process and derive complex events                               \\ \hline
\end{tabular}
\setlength{\abovecaptionskip}{0pt}
\setlength{\belowcaptionskip}{-15pt}
\caption{Description of differences in traditional \ac{ICN} vs \system architecture ("-" means no support).}
\label{tab:differences}
\end{table*}
\subsection{Unified Communication Approach} \label{subsec:unified}

In this section, we explain the extension of the \ac{CCN} data plane to enable coexistence of consumer and producer initiated communication.
In our approach, each \ac{CCN} node $n \in N$ maintains a Content Store or cache (\ac{CS}), a Pending Interest Table (\ac{PIT}), a Forwarding Information Base (\ac{FIB})
 and a \ac{CEP} engine. In the following, we explain the function of these main building blocks and the data plane handling. Subsequently, we provide details on the newly introduced packets \AddQInterest, \RmQInterest and \DataStream.

\subsubsection{Node Components}~\label{subsubsec:components}
The \ac{CS} stores all the data objects associated with the \ac{CCN} \Interest and the time-stamped data objects associated with the \AddQInterest.
The data object either returns the result of an \Interest or a CEP query $q$ in the \Interest or \AddQInterest packet. For instance, if the query is the sum of victims in a disaster location, then the data object contains a value, e.g., 20. Hence, if multiple consumers are interested in the same $q$, the query is not reprocessed but the data is fetched directly from the \ac{CS}.
To ensure that a fresh data object is returned from the cache once a \AddQInterest is registered 
and new \DataStream packet is received the query processing is triggered reactively. The reactive handling of the \DataStream ensures that
always up to-date data objects are stored in the cache, while old entries are discarded (left cf. \Cref{fig:inetcep}).

The \ac{PIT} stores the pending \Interest name with $q$ so that the \emph{complex event} ($ce$) could follow the path, i.e., the in-network state created in \ac{PIT} to the consumer.
The $\mi{face}$ information is also stored in the \ac{PIT} entry to keep track of consumers interested in $qname$. In contrast to consumer-initiated interaction, the $ce$ must be notified to the interested consumers as and when detected in real-time.
Thus, as new data is received, the \emph{qname} (name prefix of the query) in the \ac{PIT} are re-evaluated as explained later in Algorithm~\ref{algo:addqianddatastream}.
In the former, we are referring to only continuous \DataStream packet since we also support fetching discrete data from producers (this is handled similarly using a \ac{CCN} \Interest packet as explained in Section \ref{sec:preliminaries}).

The reasons why we distinguish between \AddQInterest ($qname$) and CCN \Interest packets are: (i) $qname$ is invoked on receipt of \AddQInterest as well as the \DataStream packet, (ii) removal of the \ac{PIT} entry is not based on a \Data packet retrieval but on the reception of the \RmQInterest packet, and (iii) $qname$ retrieves \Data packets asynchronously.
In summary, by asynchronously handling the $qname$ instead by 3-way message exchange and efficiently managing \ac{PIT} entries, respectively. 
We store $qname$ in \ac{PIT} and deliver complex events to the consumers when a new data object is received in the \DataStream packet. 

The \ac{FIB} table is populated as the producer multicasts to the broker network leaving a trail to the data source.
Finally, the \ac{CEP} engine holds the processing logic $f_\omega$ for each operator $\omega$ and is responsible for parsing, placement, processing, and returning the result to the next node towards consumer (cf. \Cref{subsec:QE}).

\subsubsection{Data Plane Handling}\label{subsubsec:handling}

\begin{figure}[t]
    \centering
    \includegraphics[width=\linewidth]{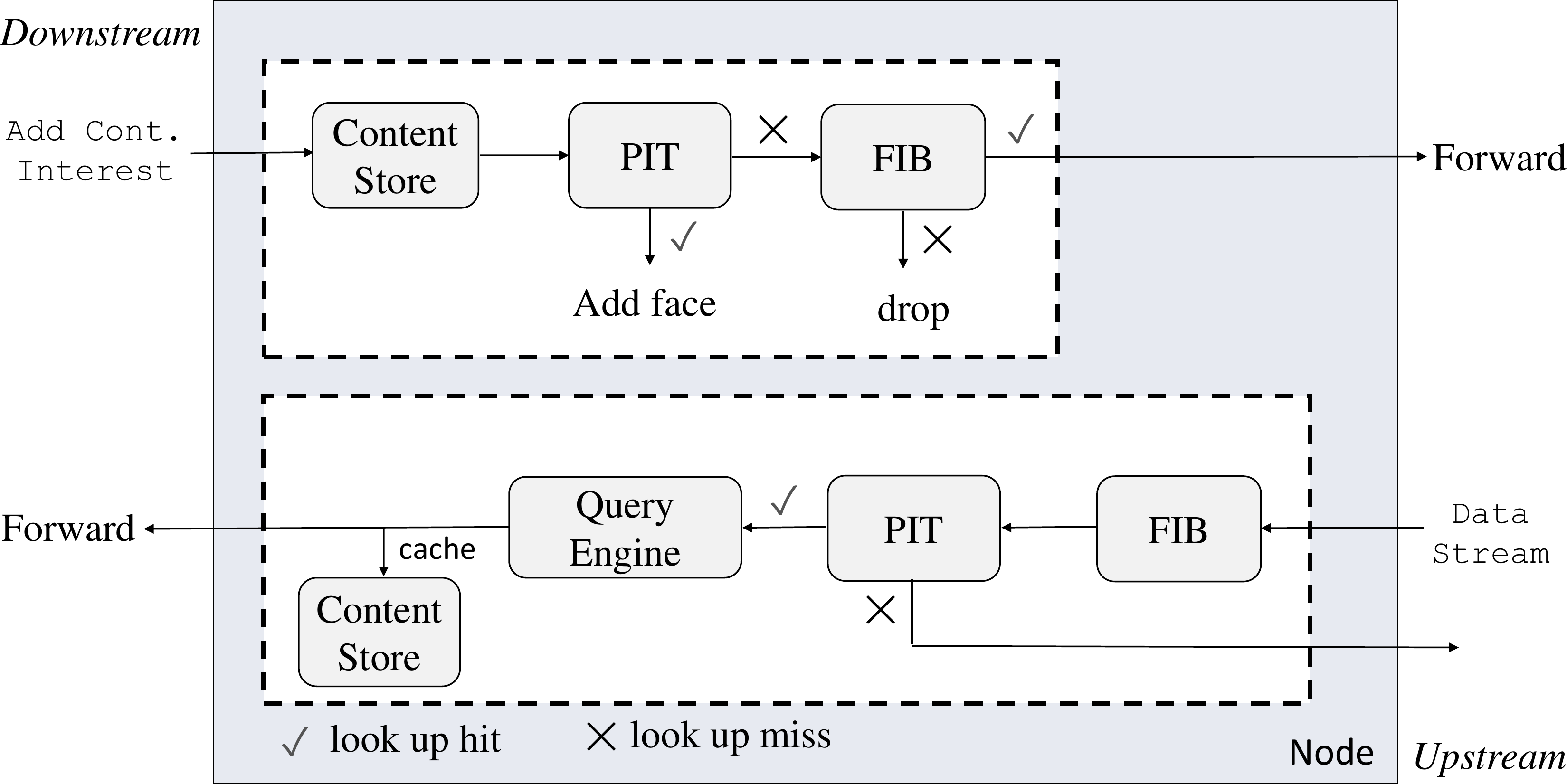}
    \setlength{\abovecaptionskip}{-5pt}
    \setlength{\belowcaptionskip}{0pt}
    \caption{High level view of packet handling in \system architecture.}
    \label{fig:overall}
\end{figure}

In Algorithm~\ref{algo:addqianddatastream} (lines~\ref{algline:addqistart}-\ref{algline:qiinputstop}) and \Cref{fig:overall}, we define the handling of \AddQInterest and \DataStream packets at the broker in a \ac{CCN} network. In \system algorithms, we express \emph{events}, \emph{packets}, and their \emph{handlers} using the definition of the asynchronous event-based composition model~\cite{Cachin2011}. Hence, events and packets are denoted as:
$\langle \textit{Event or Packet Type} | \textit{Attributes,...} \rangle$.

When an \AddQInterest arrives, the broker checks if the (up to-date) data object corresponding to the $qname$ already exists in the \ac{CS}. If this is true, the broker forwards the data object to the consumer. In addition, the broker checks the PIT for an entry of the face, in contrast to the handling of a normal \Interest packet for two reasons. (1) The data object found could correspond to the \Interest packet instead of an \AddQInterest so there should be an entry related to the \AddQInterest packet in PIT. It is important to note that the PIT entry has a flag indicating that the entry is related to \AddQInterest or \Interest packet. (2) The data object found could be from another consumer, so still the face of the new consumer needs to be added to the PIT (lines~\ref{algline:csstart}-\ref{algline:csstop}).
If a PIT entry is not found, then the broker discards the packet. This is because the $qname$ is already processed at one or more brokers.

If the cache entry is not found, the broker continues its search in the \ac{PIT} table (lines~\ref{algline:pitstart}-\ref{algline:pitstop}).
If $qname$ is found in \ac{PIT}
and the face corresponding to the query interest does not exist (lines~\ref{algline:facelist}-\ref{algline:discardinpit}), a new ${face\_id}$ (from which the interest is received) is added. Conversely, if the face entry is found in \ac{PIT}, this means the $qname$ is being processed and hence the packet is discarded (lines~\ref{algline:notfoundinpit}-\ref{algline:qiinputstop}).
However, if no entry in \ac{PIT} exists, this means that the consumer's interest reaches first time at the broker
network. Therefore, a new entry for $qname$ is created in the PIT and the \AddQInterest packet is forwarded to the face found in the FIB
(lines~\ref{algline:createpitentry}-\ref{algline:addqistop}).

A \DataStream packet is also handled similarly to the \Data packet (lines~\ref{algline:dsstart}-\ref{algline:dsstop}), except for the fact that the query processing is triggered if \DataStream satisfies a query interest in \ac{PIT} and it is a new packet  (lines~\ref{algline:checkifdsstart}-\ref{algline:checkifdsend}). This means that the $qname$ performs an operation on the data object contained in the \DataStream packet. In this case, query processing is triggered because it may contribute to the generation of a new $ce$. 
In addition, if the broker does not have a matching $qname$ in \ac{PIT} entry, this means it is not allocated to operator graph processing and hence it is forwarded to the next broker (line~\ref{algline:dsstop}). The \DataStream is forwarded if there are consumers downstream by looking at the \ac{FIB} entry.

When a \RmQInterest packet is received at a broker, the node looks up its \ac{PIT} table for an entry of the $qname$. If found, it removes the \ac{PIT} entry for $qname$ and the \RmQInterest packet is forwarded to the next node.
It is done in a similar way as the \ac{PIT} entry corresponding to a \ac{CCN} \Interest packet is removed when a matching \Data packet is found.

\IncMargin{1em}
\begin{algorithm}[t]
\scriptsize
\KwVar{$\mi{CS} \leftarrow \text{content store of current node}$ \\
$\mi{PIT} \leftarrow \text{pending interest table of current node}$ \\
$\mi{FIB} \leftarrow \text{forwarding information base}$\\
$\mi{qname} \leftarrow \text{name prefix including query}$ \\
$\mi{result} \leftarrow \text{query result}$ \\
$\mi{facelist} \leftarrow \text{list of all faces in PIT}$\\
$\mi{ds} \leftarrow \text{payload of \DataStream packet}$\\
$\mi{data} \leftarrow \text{data object that resolves the qname}$ \\
$\mi{ts} \leftarrow \text{timestamp in event tuple \DataStream} $
}
\BlankLine
\Upon{ \textbf{receipt } $\langle$ \AddQInterest $| qname \rangle$ \text{packet} }
{ \label{algline:addqistart}

	\If{ $qname$ is found in  $\mi{CS.\textsc{lookup}()}$} { \label{algline:csstart}
		$ \mi{data} \leftarrow \mi{CS}.\textsc{fetchContent}(\mi{qname}) $\;
		\If { $qname \text{ is found in } \mi{PIT.\textsc{lookup}()} $} {
		    	$\textsc{ProcessPIT}(\mi{qname}, \AddQInterest)$\;
		}
		$ \text{return } \mi{data}$\; (Discard \AddQInterest) \label{algline:csstop}
	}
	\ElseIf { $qname \text{ is found in } \mi{PIT.\textsc{lookup}()} $} { \label{algline:pitstart}
		$\textsc{ProcessPIT}(\mi{qname}, \AddQInterest)$\;		\label{algline:pitstop}
	}
	\ElseIf { $qname \text{ is found in } \mi{FIB.\textsc{lookup}()} $} { \label{algline:createpitentry}
	 	$\mi{PIT}.\textsc{addFace}(\mi{qname})$\;
	 	Forward \AddQInterest\; \label{algline:addqistop}
	}
	\Else {
	    Discard \AddQInterest \label{algline:addqistop}\;
	    }
}

\Upon{ \textbf{receipt } $\langle$ \DataStream $ | ds \rangle$ \text{packet} }
{ \label{algline:dsstart}
	\ForEach{ $\mi{qname} \in PIT$ }
	{
		\If { \texttt{ds} \text{ satisfies } $qname$ } {
			$\textsc{ProcessPIT}(\mi{qname}, \DataStream)$	\; \label{algline:processPITDS}
		}
		\Else {
			Forward \texttt{DataStream}\;  \label{algline:dsstop}
		}
	}
}
\Function{$\textsc{ProcessPIT}(\mi{qname}, \mi{packet})$}
{	\label{algline:qiinpitstart}
	\If{$\mi{packet} \text{ is } \texttt{DataStream} \text{ and } \mi{packet}.\mi{ts} > \mi{qname.ts}$} { \label{algline:checkifdsstart}  
		\KwTrigger{ $\langle \mi{deployAndProcess} | qname \rangle$ } (Refer Algo. \ref{algo:parser})\; \label{algline:processing}
		Forward $\mi{packet}$\; \label{algline:checkifdsend}
	}
	\Else {
		$\mi{facelist} \leftarrow \mi{PIT}.\textsc{getFaces}(\mi{qname}) $\; \label{algline:facelist}
		\If { $qname.\mi{face}$ is not found in $\mi{facelist}$} {  \label{algline:notfoundinpit}
			$ \mi{PIT}.\textsc{addFace}(\mi{qname}) $\; \label{algline:discardinpit}
			}
		Discard $\mi{packet}$\;	 \label{algline:qiinputstop}
	}
}

\setlength{\belowcaptionskip}{-15pt}
\caption{\AddQInterest and \DataStream packet handling.}
\label{algo:addqianddatastream}
\end{algorithm}

To summarize, in~\Cref{tab:differences}, we show the differences of the \system architecture in comparison to the standard \ac{CCN} architecture in terms of the packet types, the data plane, and the processing engine.
We show that with minimum changes in the data plane, we support both \emph{consumer-} and \emph{producer-}initiated traffic. 

\subsubsection{Rate-based Flow Control}\label{subsubsec:flowcontrol}
Flow imbalance is a well-known problem also in IP-based \emph{push} approaches. The link capacity or even the nodes processing capacity can be a bottleneck in forwarding data streams which can cause heavy packet loss and in the worst case can disrupt the system completely. CEP systems like Flink deal with this issue using a credit based flow control mechanism~\footnote{A deep dive into Flink´s network stack. \url{https://flink.apache.org/2019/06/05/flink-network-stack.html} [Accessed on 31.10.2020].}, although it has many similarities to rate-based flow control, it is often regarded as lossy~\cite{NAP1997}. Rate-based flow control mechanisms are lossless as they avoid buffer overflow by always keeping the rate below or equal to what the network can offer~\cite{Charny1995}.
In the context of \system, when continuous data streams are to be received, we propose a distributed algorithm for rate-based flow control. Each producer maintains an estimate of its optimal event rate. Initially, the producer uses its desired sending event rate as an estimate and eventually updates this estimates based on the periodic sending of management packets. The management packet contains two fields, the first field is called the \emph{underloading} bit or in short \emph{u-bit}. The second field is used to contain the next event rate estimate for the producer, known as the \emph{stamped rate}.

Before sending the management packet, the producer puts its current rate estimate in the stamped rate field, and clears the u-bit. If the demand of a flow is below the allocation it receives from the network in the management packet, the source sets the u-bit of the outgoing management packet to 1. Each CCN router / producer monitors its event traffic and calculates its available capacity per flow. This quantity is referred to as \emph{advertised rate}. When a management packet arrives, the CCN router compares the stamped rate with the advertised rate. If the stamped rate is greater or equal to the advertised rate, the stamped rate is set to the advertised rate and the u-bit is set in the management packet. Otherwise, the CCN router does not change the fields of the management packet.

When the rate control packet reaches the destination, i.e., the consumer, the stamped rate contains the minimum of the producer's rate estimate (at the time the management packet was sent) and all rates that flow is allowed to have by the CCN routers in its route. The consumer sends the management packet back to the producer. After a full round trip of the network, the setting of the u-bit indicates whether the flow is constrained along the data path. In other words, the rate is limited by some CCN router in the path and cannot be increased. In this case, the producer adjusts the stamped rate of its outgoing management packets to the stamped rate of its incoming management packet. If the u-bit is clear, the producer stamped rate is increased to the advertised rate value.

The advertised rate is computed as follows. The CCN routers maintain a list of all seen stamped rates, referred to as recorded rates. The set of flows whose recorded rate is higher than the advertised rate, are considered as unrestricted flows and are denoted as $\mathcal{U}$. Similarly, the flows with recorded rate below the advertised rate are regarded as restricted flows or $\mathcal{R}$. The flows in restricted rate set $\mathcal{R}$ are assumed to be the bottleneck links at the same or another CCN router. While the flows in the unrestricted set $\mathcal{U}$ are those for which a restricted rate has not been computed yet on this CCN router. Each CCN router, on receiving a management packet from a flow which is currently unrestricted, will compute a new stamped rate for this flow, under the assumption that this CCN router is a bottleneck for this flow. This will cause the CCN router to recompute its advertised rate as described and to insert a new rate into the stamped rate.
Given sets $\mathcal{U}$ and $\mathcal{R}$, the advertised rate ($\mu$) is calculated as the link capacity not used by flows in $\mathcal{R}$ available per flow in $\mathcal{U}$.

\begin{equation}
    \mu = \frac{\mathcal{C} - \mathcal{C}_{\mathcal{R}}}{n - n_{\mathcal{R}}}
\end{equation}

Here, $\mathcal{C}$ is the total capacity of the link, $\mathcal{C}_{\mathcal{R}}$ is the total capacity consumed by all restricted flows, $n = f + kb$ and $n_{\mathcal{R}} = f_\mathcal{R} + kb_{\mathcal{R}}$ with $f, b, f_\mathcal{R}$ and $b_{\mathcal{R}}$ being the number of total and restricted forward and feedback flows traversing the link respectively. For $k = 0$ and $k = 1$, $n$ and $n_{\mathcal{R}}$ are the total number of flows and the number and the number of restricted flows traversing the link.

Thus, by controlling the flow of events using the estimates provided by the rate-based flow control algorithm we expect timely processing of events without any event loss.

\input{sections/design-parser_wo}

%% file: sections/systemmodel_wo.tex
\subsection{Architecture Overview}

Every \ac{CCN} node can act either as a \emph{producer}, a \emph{consumer} or a \emph{broker}. A \ac{CCN} node here refers to a system entity that complies to the \ac{CCN} protocol as explained in \Cref{sec:preliminaries}. A CCN broker is an in-network element, i.e., an \system aware \ac{CCN} router, while a producer or consumer is an end device, \eg a sensor or a mobile device.
On the one hand, consumers can request a specific data item using an \Interest packet, where brokers forward the request received by consumers to support request-reply interaction, as illustrated in \Cref{fig:inetcep} on the left.
The producer replies with a data object contained in a \Data packet.
On the other hand, brokers process an unbounded and ordered \emph{data streams} generated by producers,  as illustrated in \Cref{fig:inetcep} on the right and  explained below.

A producer multicasts the data stream (\DataStream packet) towards the broker network, which disseminates the stream through the network towards consumers that expressed their interest through \AddQInterest.
A consumer issues a query by sending an \Interest or an \AddQInterest packet comprising a query $q$ (top right \Cref{fig:inetcep}) identified by a $qname$. Depending on the packet, the consumer receives the query result once or continuously. The \ac{CEP} query $q$ is processed by interconnected brokers forming a \emph{broker network}. 
The $qname$ is stored in the \ac{PIT} just like the \Interest name on the receiving broker until 
\RmQInterest packet is received that triggers the removal of $qname$ or \QInterest from the \ac{PIT} (bottom right \Cref{fig:inetcep}).

The query $q$ induces a directed acyclic operator graph,
where a vertex is an operator $\omega \in \Omega$ and an edge represents the data flow of the data stream $D$. Each operator $\omega$ dictates a processing logic $f_\omega$. In the following, we explain the unified communication model, the operator graph, and the quality of information model.

To achieve flow control, we model a flow from a CCN node defined as a uni-directional data transfer from one node to another with feedback on the opposite direction on the same route. Flows are considered to be independent of each other and they can change dynamically. Furthermore, packets in a flow are assumed to arrive in order (FIFO).

\subsubsection{Unified Communication Model}\label{subsec:comm_model}
The basic packet structure of a \ac{CCN} packet
includes a fixed-size header whose size depends on the packet type, a variable length data object name, the optional name's TLV (type length value format), some optional TLVs, and finally the payload\footnote{For more information, we refer the reader to the IETF draft \url{https://tools.ietf.org/html/draft-ccn-packet-header-00}.}.

We provide five types of packets in CCN to support both kinds of interaction patterns.
\Interest (\emph{request}) packet is equivalent to CCN's Interest packet that is used by the consumer to specify interest in the name of a data object or an operator ($\omega$).
The \Data (\emph{reply}) packet is a \ac{CCN} data packet that satisfies an Interest or \AddQInterest detailed below. This packet is similar to the \Interest with  additional payload that contains the content.
The \DataStream packet represents a continuous time series data stream of the form $<ts, a_1, \ldots, a_m>$.
Here, $ts$ is the time at which a tuple is generated and $a_i$ are the attributes of the tuple. It is similar to the \Data packet in terms of structure except the packet type and the payload comprising of the data stream.
The \AddQInterest packet represents the interest in the form of a \ac{CEP} query $q$.
The structure is similar to \Interest packet and the query is encapsulated in the name.
The \RmQInterest packet represents the \ac{CEP} query that should to be removed for the respective consumer, so that it no longer receives complex events.  
The \ac{CCN} forwarding or data plane is enhanced to handle these packets. We summarize the differences to classical \ac{ICN} architectures in \Cref{tab:differences}.

\subsubsection{Quality of Information Model}
One of our main objectives while unifying the communication strategies of ICN is to minimize the delivery response time of the complex events, while maintaining high throughput of events.
On top of that, there must not be any drop of events when delivering with short response time, which might result in wrong results.
We measure the drop of events or \textit{loss rate} in forwarding as $\frac{(\texttt{total events} - \texttt{processed events})}{\texttt{total events}}$.
Here, \texttt{total events} is the number of events that were produced by the producers and should have been received at the consumers end.
\texttt{Processed events} is the number of events that were eventually received at the consumer.

Of course,  operators like window and filter would eventually reduce the output number of events based on the semantics of the operator.
Thus, to measure the loss for complex events, we refer to the \emph{accuracy} metric.
It is defined as the proximity of the derived complex events to the true complex events as a result of false positives and false negatives.
As a consequence of event drops, the derived complex events could be false positives or false negatives.
On the one hand, a false positive (FP) is defined as a complex event that should have been derived but is not or is falsely derived, while a false negative (FN) is an complex event that is derived when it should not have been derived.
On the other hand, the true positives (TP) and true negatives (TN) are correctly derived complex events, i.e., correct load predictions and not detected, respectively.

%% file: sections/design-parser_wo.tex
\subsection{CEP Query Engine in the Data Plane} \label{subsec:QE}
In this section, we present the proposed meta query language for \ac{CEP} to specify queries in CCN data plane and the query deployment process.
We have three main design goals for the meta query language and the query parser:
\begin{inparaenum}
\item distinguishing between pull and push based traffic,
\item translating a query to an equivalent name prefix of the \ac{CCN} architecture, and
\item supporting conventional relational algebraic operators and being extensible such that additional operators can be integrated with minimum changes. This is to ensure easy integration of existing and new \ac{IoT} applications.
\end{inparaenum}
We provide the definition of \system query language in Section \ref{subsubsec:definition} and the query engine in Section \ref{subsubsec:QP}.

\subsubsection{Meta Query Language for \system} \label{subsubsec:definition}
Each operator in a query behaves differently based on the input source type, i.e., consumer- and producer-initiated interaction, which is done based on the reception of a \Data packet or a \DataStream packet, respectively.
The \Data packet is processed and returned as a data object, as conventionally done in the \ac{CCN} architecture. For instance, a \texttt{Join} operator placed on broker $b$ can join two data objects, $<lat1, long1>$ with name prefix $/node/node1$ and $<lat2, long2>$ with $/node/node2$ to produce $<lat1, long1, lat2, long2>$ with $/node/node3$.
In contrast, a \DataStream is processed and transformed either into an unbounded output stream (another \DataStream packet), or can be transformed to derive a \Data packet, containing, \eg a \texttt{boolean} variable depending on the \ac{CEP} query.

We express the standard CEP operators as explained in \Cref{sec:preliminaries} using the \system language below.

\begin{lstlisting} [caption={Selects a sliding window of tuples for 4s from gps source 1.}, label={query:window}, captionpos=b, abovecaptionskip=0pt, belowcaptionskip=-5pt]
WINDOW(GPS_S1, 4s)
\end{lstlisting}

\begin{lstlisting} [caption={Selects tuples with latitude value less than 50 pts from window Query~\ref{query:window}.}, label={query:filter}, captionpos=b, belowcaptionskip=-5pt]
FILTER(WINDOW(GPS_S1, 4s),'latitude'<50)
\end{lstlisting}

\begin{lstlisting} [caption={Performs join of the two resulting tuples from Query~\ref{query:filter} with gps source 1 and 2.}, label=query:join,captionpos=b, belowcaptionskip=-5pt]
JOIN(
  FILTER(WINDOW(GPS_S1, 4s), 'latitude'<50),
  FILTER(WINDOW(GPS_S2, 4s), 'latitude'<50),
  GPS_S1.'ts' = GPS_S2.'ts'
)
\end{lstlisting}

The stateful operators, \eg \texttt{Window} and \texttt{JOIN} must store the accumulated tuples in some form of a  readily available storage. For this, we make use of in-network cache, the \ac{CS}, that readily provides data for the window operator. This can be highly beneficial, \eg in a dynamic environment where state migration is necessary. For a complete discussion on the grammar of the language and its design, we refer the reader to our previous work~\cite{Luthra2019}.

\IncMargin{1em}
\begin{algorithm}[t]
\scriptsize
\KwVar{$query \leftarrow$ the input CEP query \\
		$\tau\mi{curList} \leftarrow$ top down list of 3 $\omega$ of tuple $\tau$ \\
		$\omega_{cur} \leftarrow$ current operator \\
		$planNode \leftarrow$ logical plan node of G \\
		$curNode \leftarrow$ current node \\
		$allNodes \leftarrow$ list of all nodes \\
		$paths \leftarrow$ list of possible paths in the topology \\
		$opPath \leftarrow$ optimal path based on the objective}
\BlankLine

\Upon{ \textbf{event} $\langle \mi{deployAndProcess} | qname \rangle$ }
{ \label{algoline:createG}
    \tcp{Query parsing}
    \If{ $query \in qname$ \text{is new}} {
    	$\tau\mi{curList} \leftarrow$  \textsc{getCurList}($query$)\;
    	$curNode \leftarrow $ \textsc{parseQuery($\tau\mi{curList})$}\; \label{algoline:stopG}
    }
    \tcp{Query placement}
    \If{$query \in qname \text{ is not placed or } query \text{ must be replaced}$} { \label{algoline:startPlacement}
        $allNodes \leftarrow \textsc{getNodeStatus} (curNode)$\; \label{algoline:getStatus}
        $paths \leftarrow \textsc{buildPaths} (curNode, allNodes)$\; \label{algoline:buildPath}
        $opPath \leftarrow \textsc{findOptimalPath} (paths, allNodes)$\;
        $\textsc{deployOperators} (optimalPath, allNodes)$\; \label{algoline:endPlacement}
        %
    }
    \tcp{Query processing}
    \ForAllP{$\omega_{cur} \in \tau\mi{curList}$} { \label{algoline:startProcessing}
        $\textsc{processOperator}(\omega_{cur}, optimalPath)$\;
    }\label{algoline:endProcessing}
}

\Function{$\textsc{parseQuery}(\tau\mi{curList})$}
{ \label{algoline:startparsing}
	$\omega_{cur} \leftarrow $\textsc{getOperator}$(\tau\mi{curList})$\;
	$\mi{nfnExp}  \leftarrow \textsc{constructNFNQuery}(\omega_{cur})$\;
	$planNode \leftarrow \text{new } \textsc{planNode}(\mi{nfnExp})$\;

	\If{$size(\tau\mi{curList}) == 1$}{
		return $node$\;
	}
	\ElseIf{$size(\tau\mi{curList}) > 1$}{

		\textsc{parseQuery}($\tau\mi{curList}.left)$\;
  		\textsc{parseQuery}($\tau\mi{curList}.right)$\;
		return $planNode$\; 	\label{algoline:stopparsing}
    }

}

\caption{Operator tree creation, placement and processing}
\label{algo:parser}
\end{algorithm}
\DecMargin{1em}

\subsubsection{Query Deployment} \label{subsubsec:QP}
Given the constructs of the query language, we now explain how the queries are processed in the CCN data plane. In \system, the queries can be processed in a centralized or a distributed fashion for better scalability. For instance, when the amount of resources required to process the queries exceed (that can easily happen in case of stateful operators like joins). Moreover, the core in-network operations should not be disturbed by the execution of CEP operators.

As shown in \Cref{fig:overall}, \AddQInterest and \DataStream packets invokes the query engine and hence parsing, placement and processing of the queries in the form of an operator graph. In Algorithm~\ref{algo:parser}, we show this process. The query parser parses and translates the query into an executable form. Furthermore, it verifies the correctness of the query by validating the parameters. For example, a join can be on a set of data streams and not a single data stream, so it needs atleast two parameters.  When a new query encapsulated in the name field of \AddQInterest packet is received on a node, the recursive function \textsc{parseQuery} performs the aforementioned checks (lines~\ref{algoline:createG}-\ref{algoline:stopG}). It is important to note that this function is invoked only when a new query is received. This way, we provide query reuse, an important functionality of a CEP system needed when same query is invoked from multiple consumers~\cite{Ottenwalder2014b}. 

We express the \system query parser as a recursive algorithm to map the query  in order to generate an equivalent \ac{NFN}'s \emph{ $\lambda$ expression} which is processed as follows.
A CEP query is transformed into an operator graph $G$ (lines~\ref{algoline:createG}-\ref{algoline:stopG}), which is an acyclic tree  defined as a tuple $\tau = (L, S, R)$. Here, $L$ (left sub tree) and $R$ (right sub tree) are binary trees or an empty set and $S$ is a singleton set, \eg a single operator ($\omega$). The query parser starts parsing the query in a specific order, i.e., in a top-down fashion that marks the dependency of operators as well. This implies each leaf operator is dependent on its parent. Thus, the parser starts by iterating top down the binary tree starting from the root operator $\omega_{cur}$ (line~\ref{algoline:stopG}), where $cur = root$ in the first step. The traversal is performed in a depth-first pre-order manner (visit parent first, then left (L) and then right (R) children) (lines~\ref{algoline:startparsing}-\ref{algoline:stopparsing}).

To determine the query placement, we make use of \Interest packets with a flag indicating that they are for management (or internal) use. Furthermore, we make use of a node status service that is probed to get the information on the nodes as well as the links between them. This can run, for example, on a monitoring server that collects information on the network. This information includes distinct quality of service (QoS) metrics widely used for operator placement~\cite{Luthra2018} such as response times, bandwidth, energy information (crucial for IoT), and finally, we also include the rate estimate from the rate-based flow control mechanism (discussed in \Cref{subsubsec:flowcontrol}). The response $r_p$ is encapsulated in the payload of \Data packet and is of the form $start|b_i|end$ = $r_p|{b}_p|{qos}_{p_i}|\mu_p$, where $start$ is the originating node of the path, $b_i$ are the intermediate nodes or the brokers and $end$ is the end node of the path. On the right hand side, we have the end to end response time (${r}_p$), bandwidth (${b}_p$), $qos_{p_i}$ for other QoS metrics  and $\mu_p$ is the rate estimate of this flow in the path $p$.

Hence, when a \AddQInterest encapsulating a query is received on a node, it probes the decentralized service for its neighbor nodes status by means of an \Interest packet with name-prefix \emph{node/nodeStatus}. The node status service sends the information $r_p$ for each path including its neighboring nodes. The placement coordinator (receiver of the query) then depicts the optimal path where the query $q$ is deployed (Line~\ref{algoline:startPlacement}--\ref{algoline:endPlacement}). It is important to note that the query placement is trigerred in the beginning for each query and when a replacement request is received, for example, due to the dynamic changes in the IoT network. This further accelerates the query processing in the data plane.

We have now the formulated operator graph and the way it is placed on the distributed nodes. However, the data stream still only reaches the node where the query was triggered. The data stream and the intermediate events must also flow down the operator graph to derive the complex event. For this purpose, the so-called \emph{sub} continuous interests are sent down the optimal data path such that the pending interest tables are updated for all nodes on the path. Finally, the operator graph is processed in an asynchronous parallel manner as seen in line~\ref{algoline:startProcessing}--\ref{algoline:endProcessing}.

%% file: sections/evaluation.tex
\begin{table}[t]
\begin{center}
\small
\begin{tabular}{ p{3.5cm} p{4.8cm} }
\hline
Experiment time $t_s$ & 20 min \\
Warmup time $t_w$ & 60 s \\
Number of runs & 10 \\
Topologies & \underline{Manhattan graph (WSN)}, line and tree (edge-cloud) topology \\
Number of producers & {\underline{1}, 2}  \\
Number of consumers & 1 \\
Number of brokers & \underline{5} (Manhattan), 3 (line), 8 (tree) \\
Queries & \underline{Window}, Filter, Join, Heatmap, Predict (cf Query~\ref{query:window}--\ref{query:predict}) \\
Number of cloud instances for distributed setup & \underline{1 GC c2-standard-8} and 7 AWS c5d.2xlarge instances \\
Input event rate & \underline{1}, {1000}, 10,000 and 50,000 (Poisson distribution) \\
\hline
ICN and CEP systems & INetCEP\_UCL (ours), {Apache Flink}~\cite{Carbone2015ApacheFS}, \underline{Periodic Request (PR) }~\cite{Shang2016}\\
\hline
\end{tabular}
\setlength{\abovecaptionskip}{0pt}
\setlength{\belowcaptionskip}{-10pt}
\caption{Configuration parameters for the evaluation. Default or commonly used parameters are underlined.}
\label{tab:config-parameters}
\end{center}
\end{table}

\begin{figure*}
    \centering
    \subcaptionbox{Single node with producer, consumer, and broker communicating via unified communication layer. The metrics are annotated as they are measured.}
    {\includegraphics[width=0.35\linewidth]{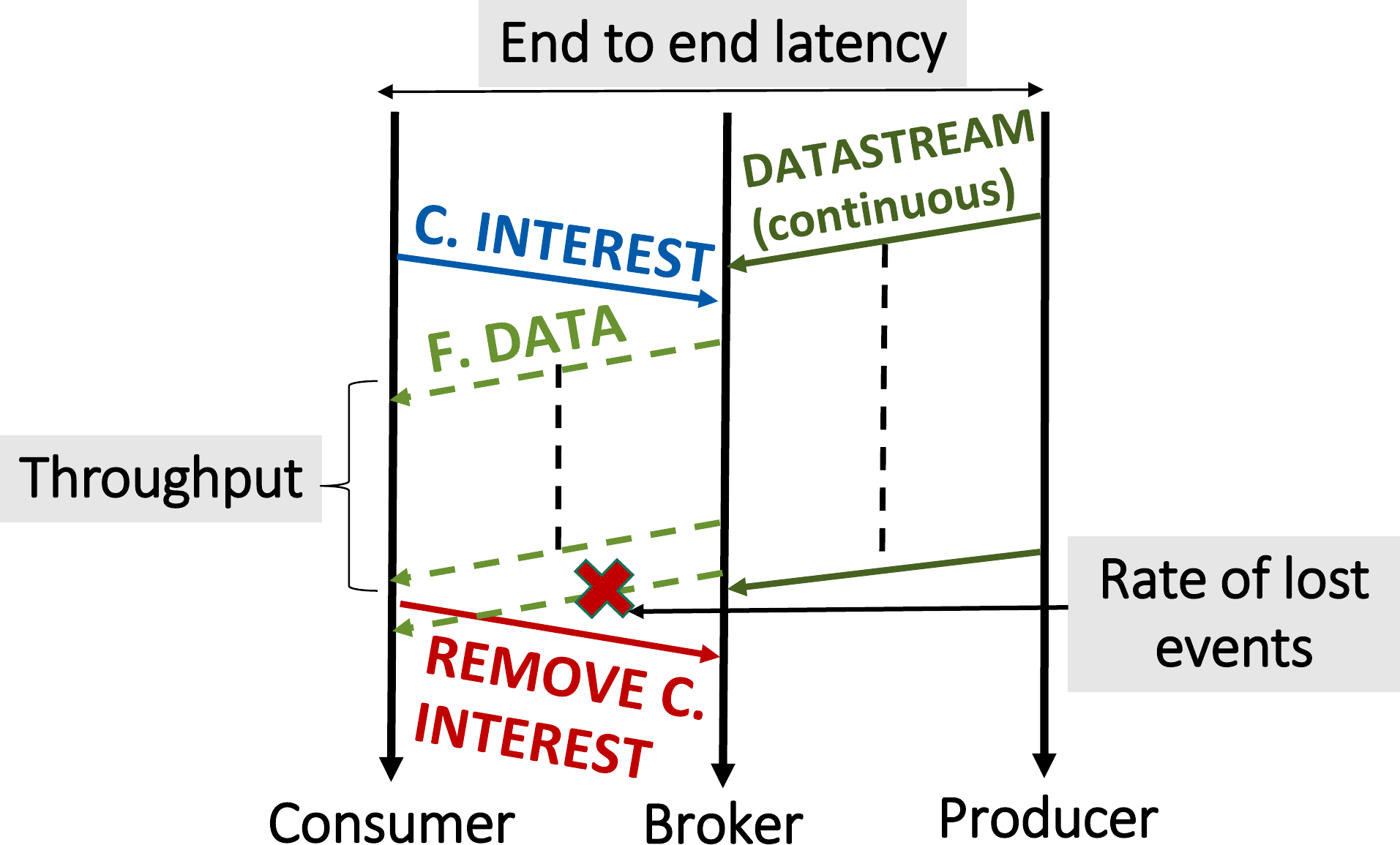}}
    \hfill
    \subcaptionbox{The different topologies Manhattan graph, line, and tree topology representing the edge-cloud network (left to right).}{
    \includegraphics[width=0.25\linewidth]{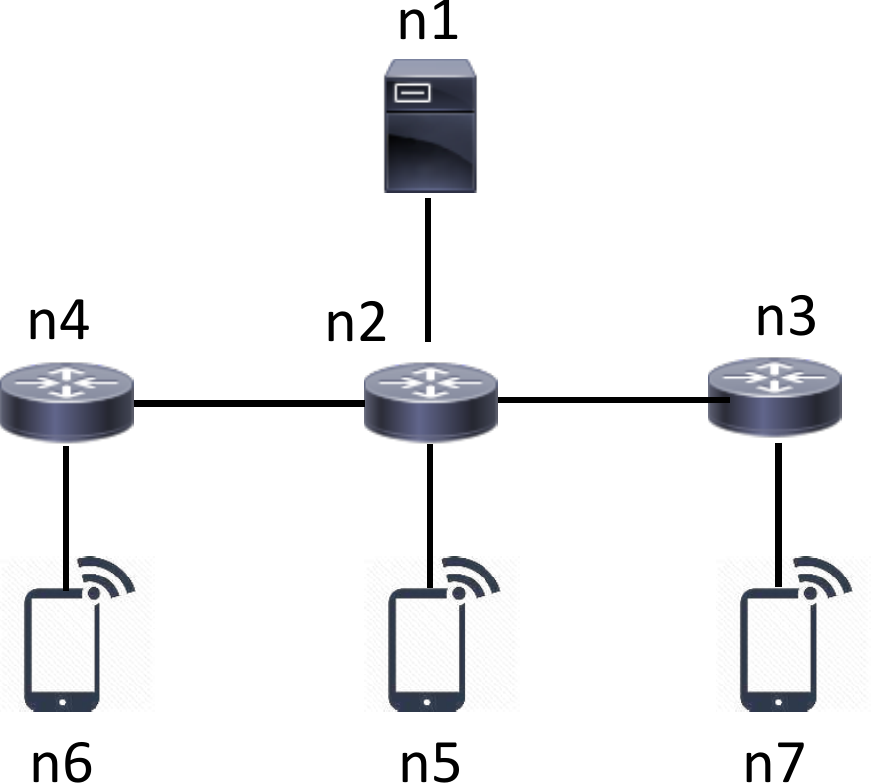}
    \includegraphics[width=0.35\linewidth]{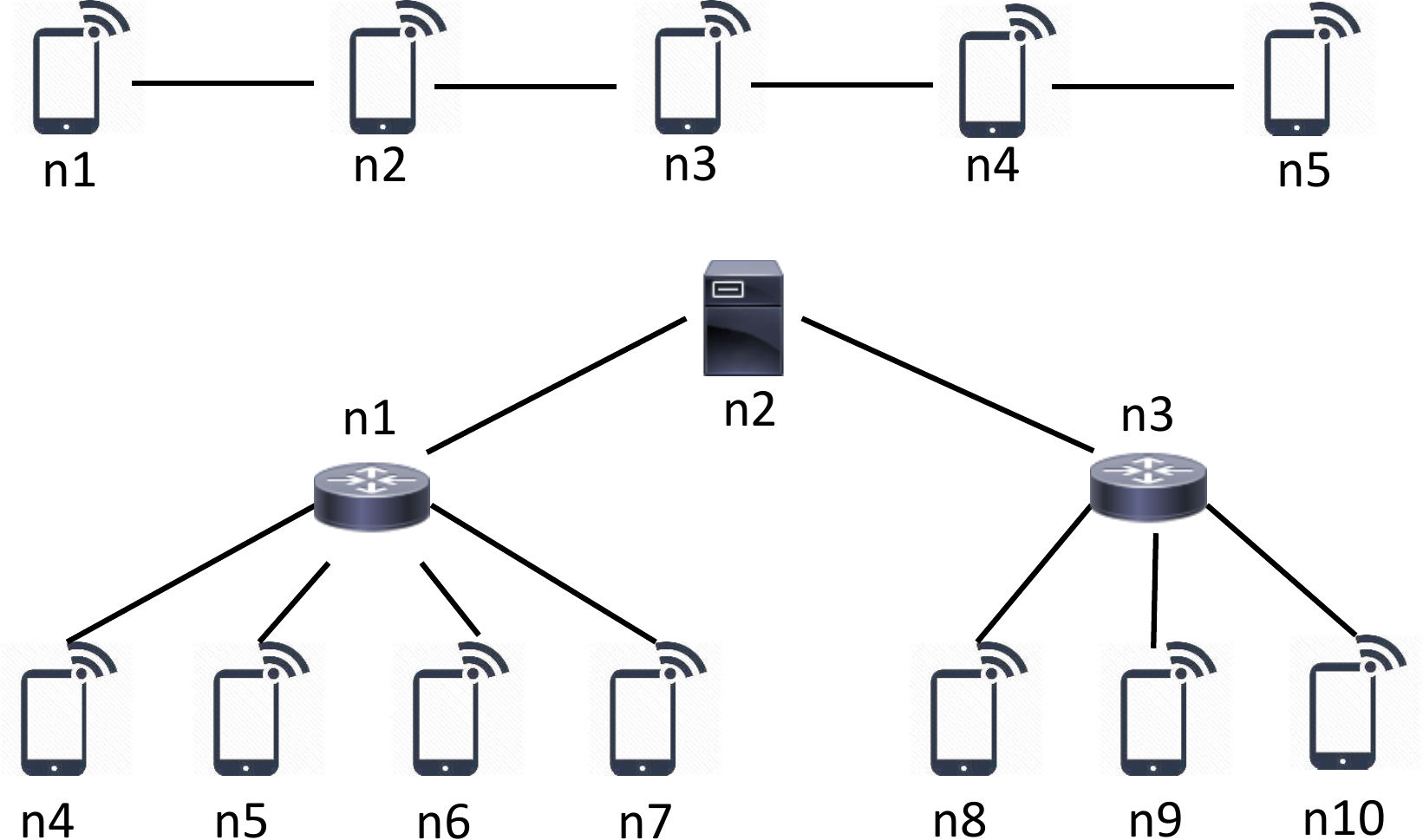}}
    \setlength{\abovecaptionskip}{0pt}
    \setlength{\belowcaptionskip}{-15pt}
    \caption{Single and distributed nodes setup.}
    \label{fig:topo}
\end{figure*}
\section{Evaluation} \label{sec:eval}

We evaluate  \system  by answering three questions:
\newline

\textbf{EQ1:} Can the unified communication approach support both consumer- and producer-initiated communication and match the performance of state-of-the-art CEP systems?

\textbf{EQ2:} What is the performance of \system when the queries are processed on a single node?

\textbf{EQ3:} What is the performance of \system when the queries are processed on distributed nodes in an edge / data-center topology?

To this end, we explain the evaluation setup in Section \ref{subsec:setup}, and answer \textbf{EQ1} in Section \ref{subsec:eq1}, \textbf{EQ2} in \Cref{subsec:eq2}, and \textbf{EQ3} in \Cref{subsec:eq3}.

\subsection{Evaluation Environment} \label{subsec:setup}

\textbf{Implementation: }
We selected the \ac{NFN} architecture~\cite{Tschudin2014} to implement our solution, due to its built-in support of resolving \emph{named functions} as so-called $\lambda$ expressions on top of the \ac{ICN} substrate.
However, currently \ac{NFN}'s communication plane is purely consumer-initiated and there is no support for CEP.
To this end, we provide a unified communication layer for co-existing consumer- and producer-initiated interactions, while doing \ac{CEP} operations in the network, on top of NFN.
Therefore, we embedded \ac{CEP} operators as \emph{named functions} while leveraging \ac{NFN}'s abstract machine to resolve them. NFN works together with CCN-lite~\cite{CCN-lite2019}, which is a lightweight implementation of CCNx and the NDN protocol. We developed unified interfaces of our design on top of NFN (v0.2.1) and CCN-lite (v2.0.1) for the Linux platform~\cite{INetCEPGithub2019}. We implemented  13,454 LOCs and 19,491 LOCs in CCN-lite and NFN-Scala, respectively.

We enhanced the NDN protocol implementation in the CCN-lite and the NFN architecture by
\begin{inparaenum}
\item including the additional packet types and their handling, as described in \Cref{subsec:unified},
\item implementing the extensible general CEP query language, parser, and CEP operators as \ac{NFN} services, as described in Section \ref{subsec:QE}, and
\item implementing the parsing, placement and processing of operator graphs in the data plane of CCN (cf. Section \ref{subsubsec:QP}).
\end{inparaenum}

\textbf{Evaluation Platform: }
We evaluated our implementation both using CCN-lite as well as the \textit{Common Open Research Emulator} (CORE) network emulator
and MACI~\cite{frommgen2018don} for distributed node evaluation in the cloud. We used the CORE emulator in order to include different topologies like line and tree.
We used virtual machines from Google Cloud and Amazon Web Services (AWS) to introduce heterogeneity in consistent to the IoT scenario. In Google Cloud we used the c2-standard-8 machine with 8vCPUs, processing speed of upto 3.8 Ghz,
and 32 GiB memory.
Using AWS we chose instances of the type c5d.2xlarge with 8 vCPUs, processing speed of upto 3.5 Ghz,
and 16 GiB memory. All the instances run a Ubuntu 18.04 server.
Here, each VM is a CCN-NFN node, which hosts a \ac{NFN} compute server encapsulating the CEP operator logic. To run the experiments, we first used a \emph{Manhattan graph}, a well-known WSN topology~\cite{Bonfils2004}, and a tree topology resembling edge and data-center networks with varying number of nodes as illustrated in~\Cref{fig:topo}. We deployed the \system architecture that executes on the NFN compute server, the CCN-NFN relay, as well as on the links. Here, as intended, the nodes communicate using the NDN protocol instead of IP.

\textbf{Datasets.} We extended the \system architecture to the two representative \ac{IoT} introduced in \Cref{sec:problem}, namely the load prediction query and the heat map query.
We extended the \system query language and \ac{CEP} operators to include the heat map~\cite{Koepp2014} and prediction operators~\cite{Martin2014} by making a few additions to our implementation in extensible query language and parser~\cite{Luthra2019}.

\textbf{Dataset 1.} For the heat map query, we use a dataset~\cite{Alvarez2018} of a field test mimicking a post-disaster situation.
The collected dataset consists of sensor data, \eg location coordinates.
Each sensor data stream has a schema specifying the name of the attributes, \eg the GPS data stream has the following schema:
\\
\begin{footnotesize}
$<ts, s\_id, latitude, longitude, altitude, accuracy, distance, speed>$
\end{footnotesize}

\textbf{Query.} We use the $latitude$ and $longitude$ attributes of this schema to generate the heat map distribution of the survivors from the disaster field scenario. A typical heat map application joins the GPS data stream from a given set of survivors, derives the area by finding minimum and maximum latitude and longitude values, and visualizes the heat map distribution of the location of the survivors in this area. For simplicity, we consider one data stream from one survivor.
Here, \texttt{GPS\_S1} is the producer of GPS sensor, \textbf{\texttt{HEATMAP}} operator represents the heat map generation  algorithm~\cite{Koepp2014}.

\begin{figure*}[t!]
\centering
\subcaptionbox{ $\mean{x}(Flink)=792.19 \mu s$, $\mean{x}(PR)= 41.82 \mu s$,  $\mean{x}(UCL)= 38.77 \mu s$.
}
{\includegraphics[width=0.3\linewidth]{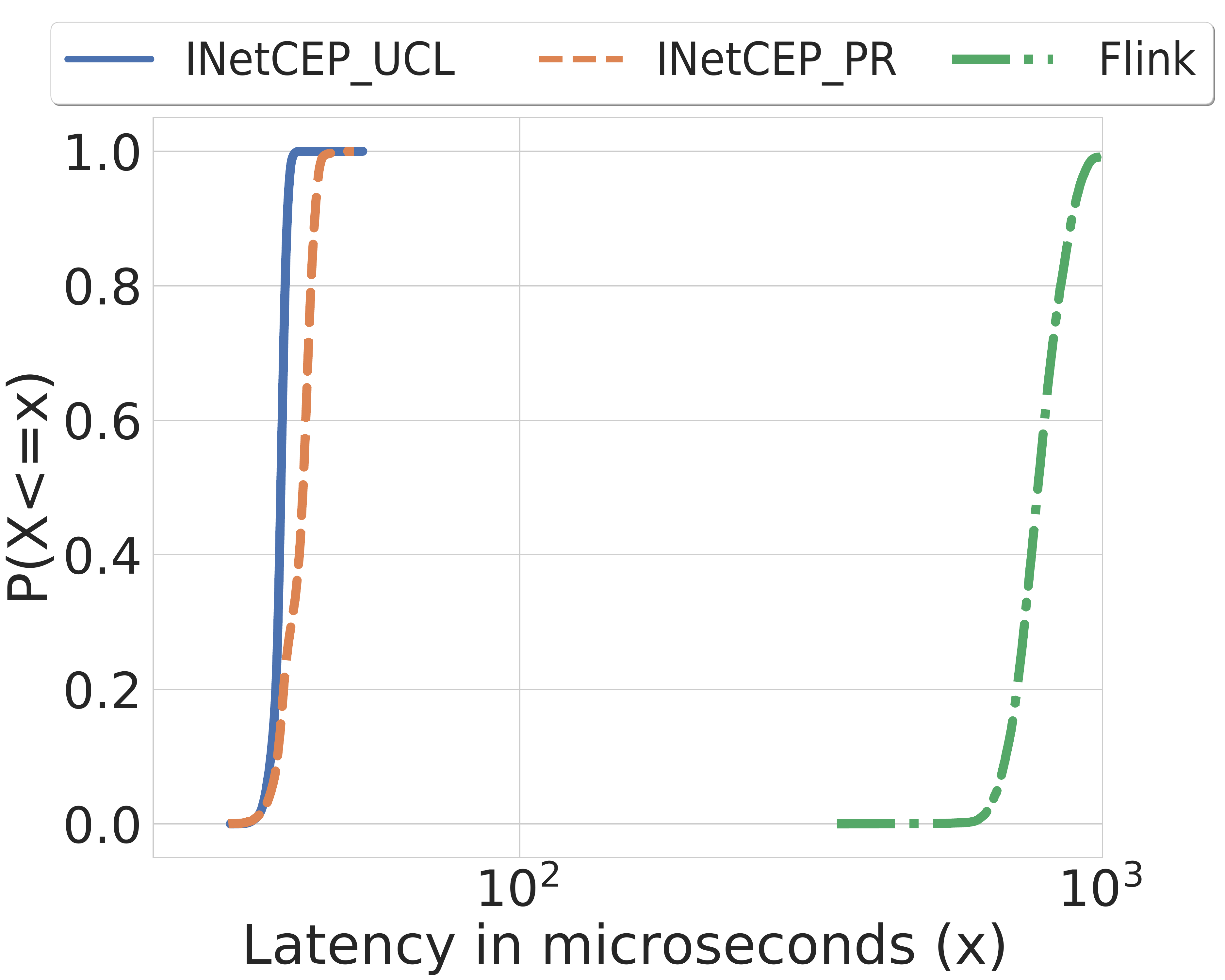}}
\hfill
\subcaptionbox{$\mean{x}(Flink)=982.16 \mu s$, $\mean{x}(PR)= 39.06 \mu s$,  $\mean{x}(UCL)= 38.20 \mu s$.
}
{\includegraphics[width=0.3\linewidth]{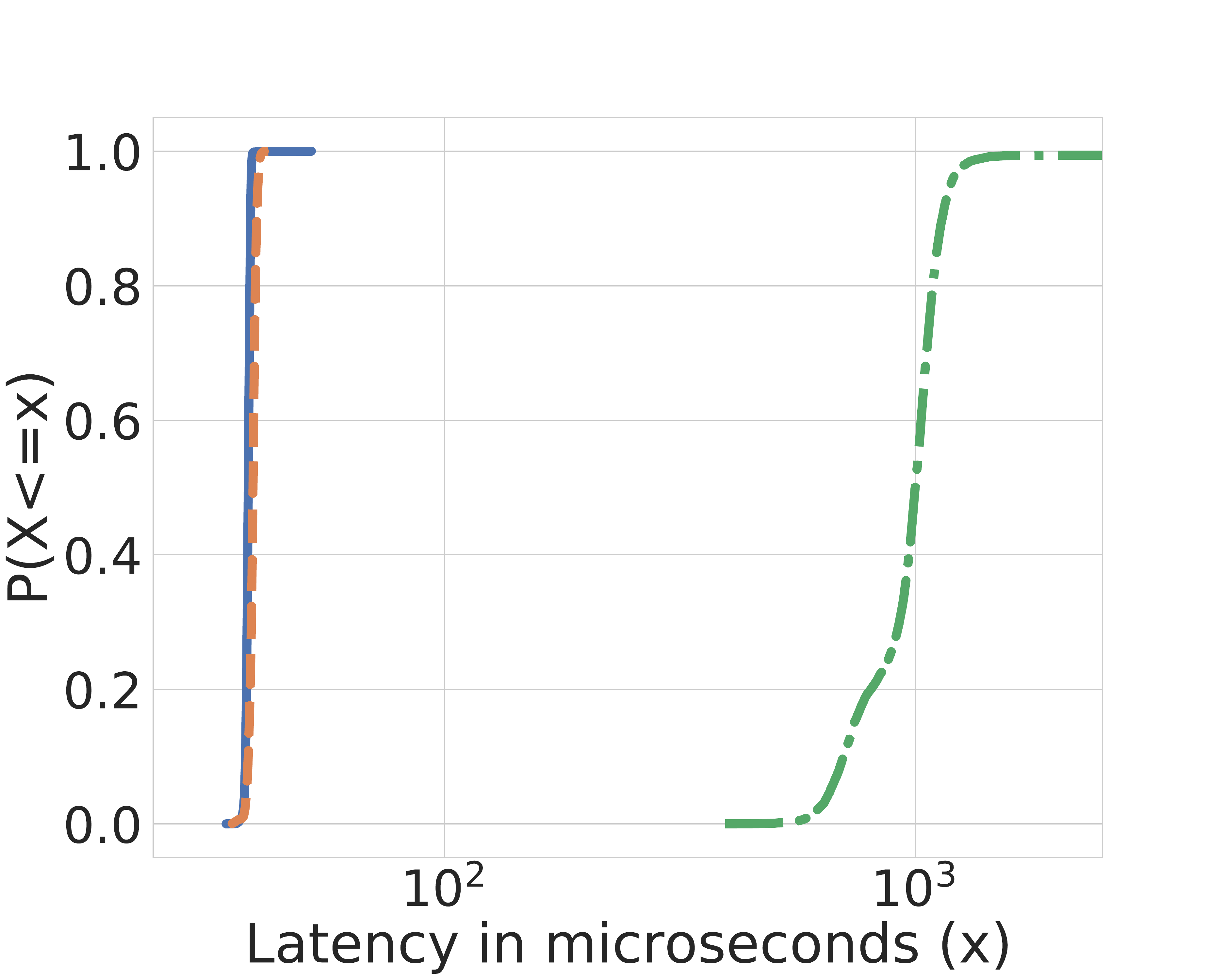}}
\hfill
\subcaptionbox{$\mean{x}(Flink)=1162.53 \mu s$, $\mean{x}(PR)= 50.38 \mu s$,  $\mean{x}(UCL)= 73.57 \mu s$.
}
{\includegraphics[width=0.3\linewidth]{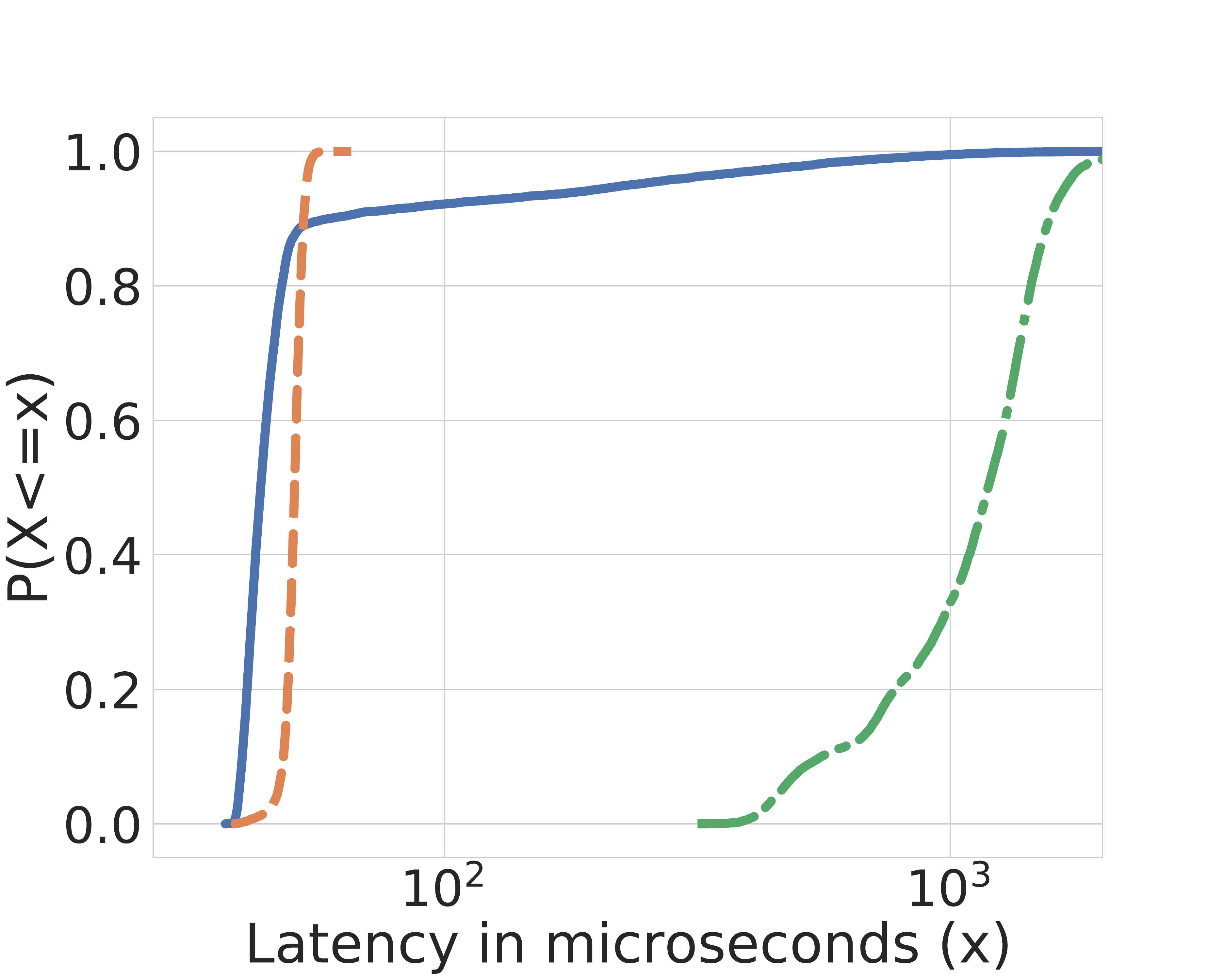}}
\setlength{\abovecaptionskip}{5pt}
\setlength{\belowcaptionskip}{-15pt}
\caption{Cumulative distribution function of end-to-end delay for Unified Communication (UCL), Flink and Periodic Request (PR) approach. Our approach is $20, 25$ and $15 \times$ faster in forwarding than Flink and mostly better than the flooding (PR) approach for an event rate of 1000, 10,000 and 50,000 events per second (left to right), respectively.}
\label{fig:evaldelay}
\end{figure*}

\begin{lstlisting}[caption={Display the heat map distribution of GPS source 1 in the given area with a given cell size.}, captionpos=b, label=query:heatmap]
HEATMAP(
  'cell_size', 'area',
  WINDOW(GPS_S1, 4s)
)
\end{lstlisting}

\textbf{Dataset 2.} The second dataset comes from the 2014 DEBS Grand Challenge~\cite{DEBS2014} scenario focused on solving a short-term load forecasting problem in a smart grid. The data for the challenge is based on real-world profiles collected from smart home installations.
The dataset captures load measurements from 
unique smart plugs with the following schema:
\\
\begin{footnotesize}
$<ts, id, value, property, plug\_id, household\_id, house\_id>$
\end{footnotesize}

\textbf{Query.} We apply an existing solution~\cite{Martin2014} to perform prediction by extending the \system architecture for smart plugs. In this query, a window of 4s of plug sensor load is analysed to generate predictions every 30 seconds.

\begin{lstlisting}[caption={Performing a prediction every 30 seconds for 1 minute into the future on the load observed by plug src 1.}, captionpos=b, label=query:predict]
PREDICT(30s, WINDOW(PLUG_S1, 4s))
\end{lstlisting}

\textbf{Metrics.} The main objective of \system  is to minimize response time, better known as \emph{end-to-end latency} in the delivery of events, while maintaining a high throughput. The latency is measured as the total time taken from the generation of the event at the producer until the complex event reaches the consumer. The second metric is \emph{throughput} which is measured as the number of events received at the consumer compared to the number of produced events. In an ideal case, the consumer should receive all the events sent by the producer given that it expresses a continuous interest on all the events from the producer. Another metric is the loss of events while forwarding or retrieving the complex event. The \emph{loss rate (\%)} is measured as $\frac{(\text{total events} - \text{processed events})}{\text{total events}} \times 100$. Finally, we are interested in the accuracy of the retrieved complex event as a consequence of loss of primary events. In \Cref{fig:topo}a, we have illustrated the metrics for a single node setup.

\textbf{Baseline. } We consider Apache Flink~\cite{Carbone2015ApacheFS}, a state-of-the-art CEP system, which so far has the best performance in terms of latency and throughput in CEP systems, to be a valid baseline. Flink consumes the continuous data stream from Apache Kafka, which acts as a data producer. Furthermore, we implemented a periodic request approach (denoted as INetCEP\_PR) equivalent to the implementation of other work ~\cite{Shang2016} in \system and compare our solution against it.
A summary of the used parameters is provided in \Cref{tab:config-parameters}.

\subsection{Performance of the Unified Communication Approach} \label{subsec:eq1}
To understand the performance of the unified communication approach (denoted as INetCEP\_UCL), we measure the metrics end-to-end latency, throughput, and loss in forwarding the events.
We consider a single node setup as shown in \Cref{fig:topo}a, where producer, consumer and broker coexists on the same node. The producer generates a continuous data stream of a GPS sensor from dataset 1 (above section). For this evaluation, we used one Google Cloud c2 instance. The consumer places a continuous interest by sending a \AddQInterest packet to the broker and starts receiving the data stream from the broker. Each evaluation run is for 20 minutes, after which the consumer removes the continuous interest by sending a \RmQInterest packet. We repeat the experiments 10 times to collect 12,029 data points for the metrics end-to-end latency, throughput, and loss rate.
\newline

\begin{figure*}[t!]
\centering
\subcaptionbox{$\mean{x}(Flink)=999.86 ev/s$, $\mean{x}(PR)= 999.63 ev/s$,  $\mean{x}(UCL)= 1000.0008 ev/s$
}
{\includegraphics[width=0.3\linewidth]{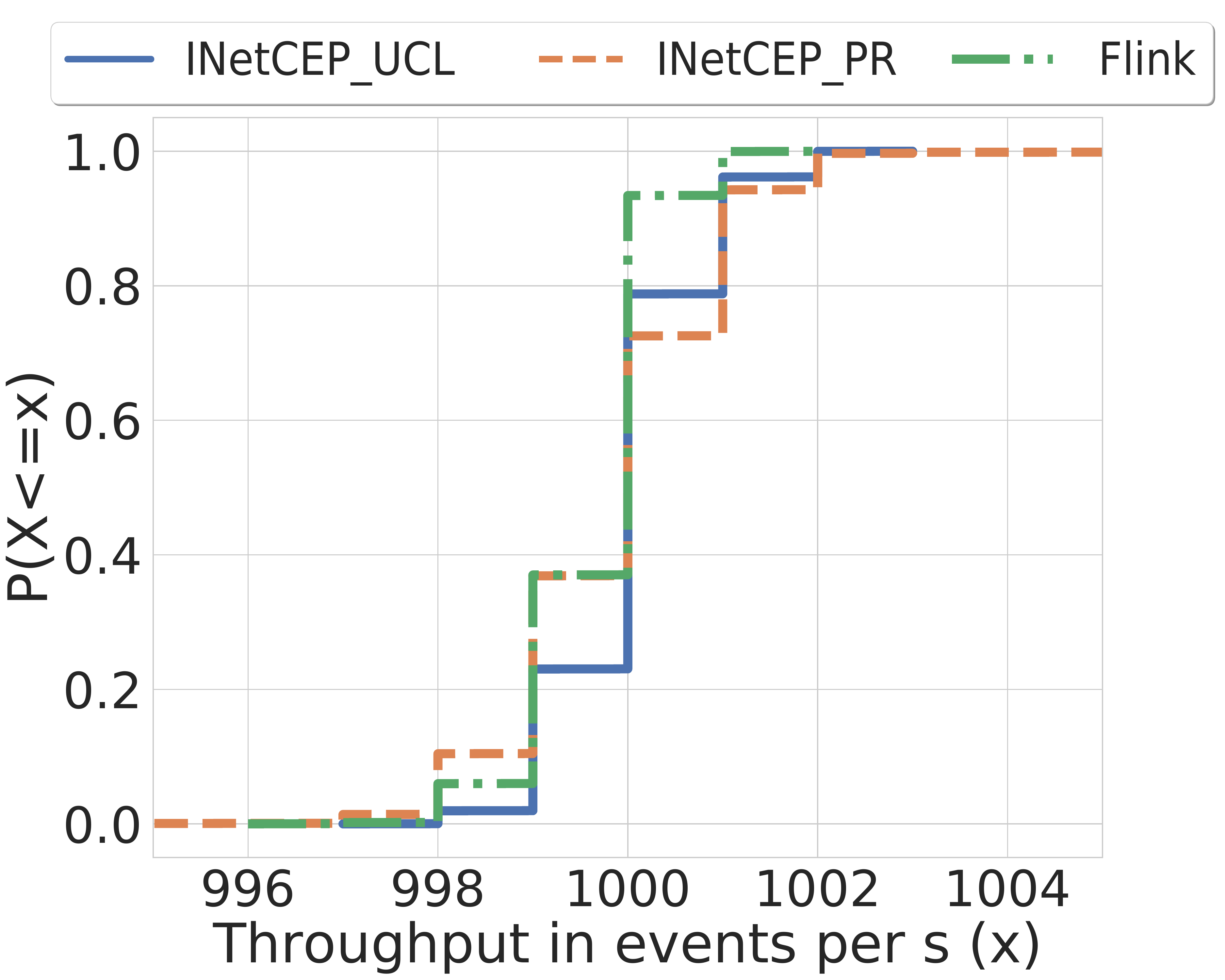}}
\hfill
\subcaptionbox{$\mean{x}(Flink)=9997.99 ev/s$, $\mean{x}(PR)= 9999.65 ev/s$,  $\mean{x}(UCL)= 10000.0008 ev/s$
}
{\includegraphics[width=0.3\linewidth]{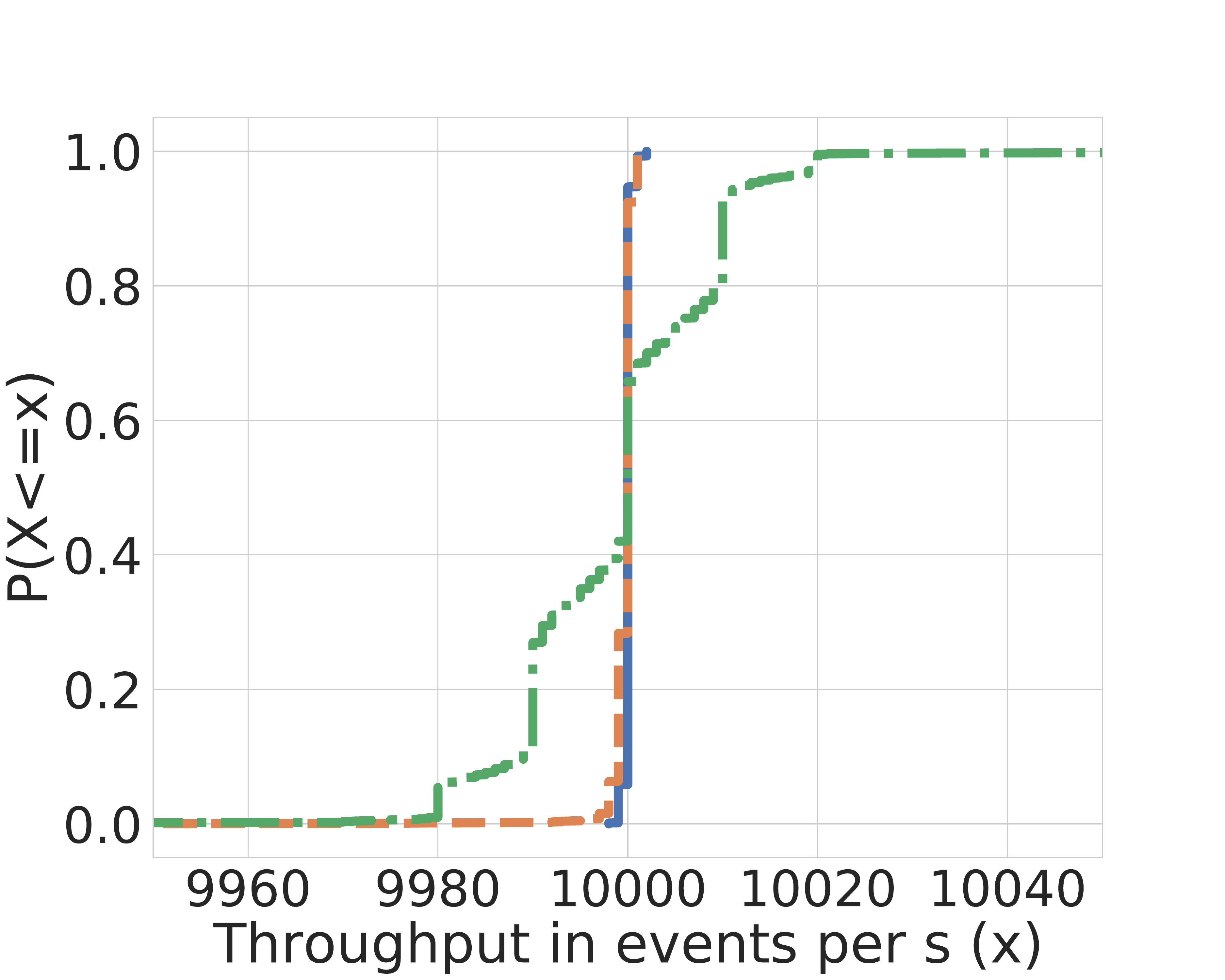}}
\hfill
\subcaptionbox{$\mean{x}(Flink)=50035.28 ev/s$, $\mean{x}(PR)= 19372.90 ev/s$,  $\mean{x}(UCL)= 49972.34 ev/s$
}
{\includegraphics[width=0.3\linewidth]{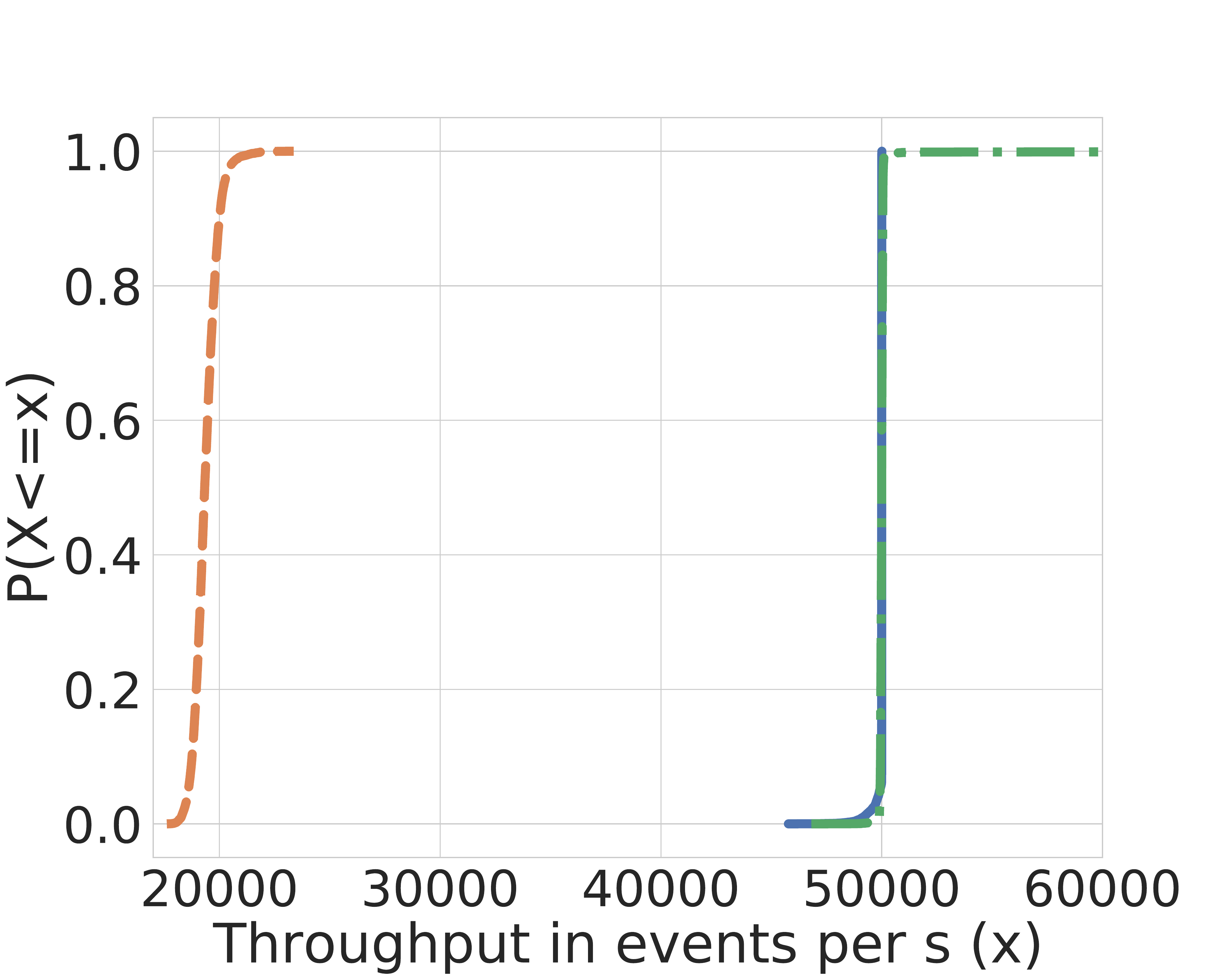}}
\setlength{\abovecaptionskip}{5pt}
\setlength{\belowcaptionskip}{0pt}
\caption{Cumulative distribution function of throughput for Unified Communication (UCL), Flink, and Periodic Request (PR). Our approach is performing equally in terms of throughput in forwarding events compared to Flink and better than the flooding (PR) approach for an event rate of 1000, 10,000 and 50,000 events per second (left to right), respectively.}
\label{fig:evaltp}
\end{figure*}

\begin{table*}[]
\scriptsize
\begingroup
\setlength{\tabcolsep}{5pt} 
\begin{tabular}{l|l|l|l|l|l|l|l|l|l|l|l|l|}
\cline{2-13}
                                            & \multicolumn{4}{c|}{\textbf{1000 events per second}}                               & \multicolumn{4}{c|}{\textbf{10,000 events per second}}                             & \multicolumn{4}{c|}{\textbf{50,000 events per second}}                             \\ \hline
\multicolumn{1}{|l|}{\textbf{System}}       & \textit{mean} &
\textit{max} & \textit{min} &
\textit{quantiles (90, 95, 99)} & \textit{mean} &
\textit{max} & \textit{min} &
\textit{quantiles (90, 95, 99)} & \textit{mean} &
\textit{max} & \textit{min} &
\textit{quantiles (90, 95, 99)} \\ \hline
\multicolumn{1}{|l|}{\textbf{Flink}}        & 0.013        &
0.014        & 0.013       &
0.013, 0.014, 0.014           & 0.135         &
0.271       & 0.069       &
0.211, 0.241, 0.265           & 0.0135        &
0.020       & 0.003       &
0.016, 0.018, 0.019          \\ \hline
\multicolumn{1}{|l|}{\textbf{INetCEP\_PR}}  & 0.038        &
0.044       & 0.036       &
0.04, 0.042, 0.043           & 0.003        &
0.004       & 0.001       &
0.004, 0.004, 0.004           & 61.185        &
61.226       & 61.153       &
61.216, 61.221, 61.225          \\ \hline
\end{tabular}
\endgroup
\caption{Packet loss rate (in \%) mean, min, max and quantiles (90, 95, 99) in forwarding 1000, 10,000 and 50,000 events per second in Flink and PR. \system do not have suffer any loss thanks to the rate-based flow control and in-network caching mechanism of INetCEP in ICN.}
\label{table:plr}
\end{table*}
\subsubsection{Forwarding Latency}
In \Cref{fig:evaldelay}, we show the cumulative probability distribution function (CDF) of the observed forwarding latency between sending at the producer and receiving at the consumer. As expected, the forwarding of the data stream that happens in the unified communication layer on the data plane is very fast with a mean delay of $38.77 \mu s$,  $38.2 \mu s$ and $73.57 \mu s$ for an event workload of 1000, 10,000 and 50,000 events per second, respectively (cf. \Cref{fig:evaldelay}a, b and c left to right). With our approach, we are $20, 25$ and $25\times$ better (in terms of mean delay) for an event rate of 1000, 10,000 and 50,000 events per second, respectively, than the baseline CEP system Flink due to several reasons. (1) Flink relies on Apache Kafka as a streaming system or producer, which inherently uses a \emph{pull-}based approach. Although it applies optimizations like aggressive batching of events to be forwarded to the consumer, it still falls short in forwarding latency because of too many requests per event. (2) Flink operates in a middleware layer, while we benefit from line rate speed of the data plane of CCN.

Furthermore, the periodic request (PR) approach, which is a \emph{pull-}based implementation on top of CCN-lite, inspired by Shang et al.~\cite{Shang2016}, performs equally well in terms of latency in receiving the data streams, but on the cost of event/packet loss, as explained later.

\subsubsection{Forwarding Throughput}
In \Cref{fig:evaltp}, we show the CDF of the observed forwarding throughput for 1000, 10,000, and 50,000 dispatched events per second, respectively (left to right), from producer to consumer. We use a single node setup as shown in \Cref{fig:topo}a.
As shown in the first plot (a),  the three approaches perform equally well. A step-wise progress of the distribution function results from the discrete nature of the metric.
In the second plot, we observe that both UCL and PR are near to the optimal throughput of 10,000 events. However, for Flink we see a divergence from the optimum. This can be explained by the batching mechanism of Flink and Kafka that forwards the events in batches such that the events do not arrive in a continuous flow but more in a batch form, which correlates with the observation of this plot.

Finally, the third plot shows the throughput results for an event rate of 50,000. We see that for the PR approach there is a heavy drop of events. We explain it as follows. The PR approach requires twice as many messages to receive the continuous data stream. Thus, for this plot to retrieve 50,000 events per second, it will send 50,000 \Interest packets. This simply floods the network and eventually we see packet drops, when the UDP buffer is full. This is not happening in Flink or in UCL, since UCL requires only a single \AddQInterest packet to retrieve the continuous stream and uses a rate-based flow control mechanism. Flink also manages the flow using a credit-based flow control mechanism.

\begin{figure*}[t]
\centering
\includegraphics[width=0.79\linewidth]{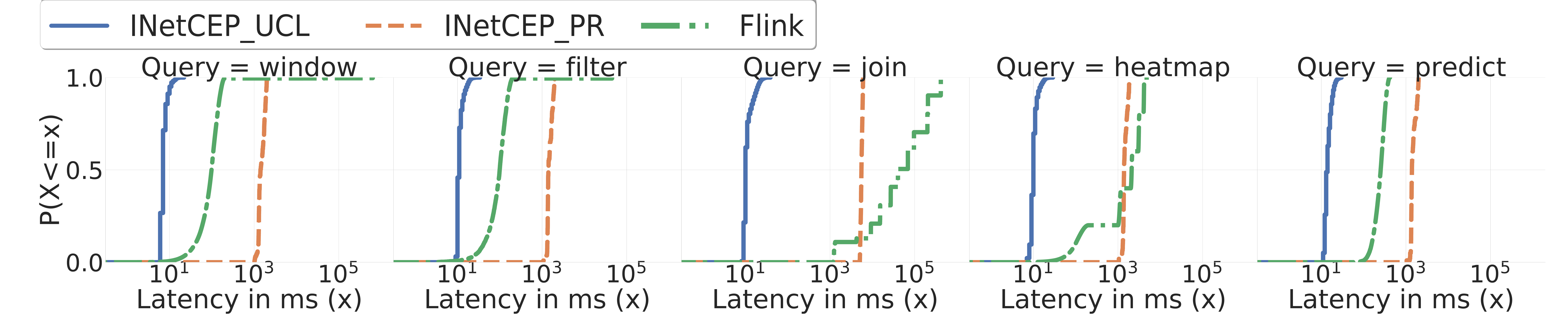}
\setlength{\abovecaptionskip}{5pt}
\setlength{\belowcaptionskip}{0pt}
\caption{Cumulative distribution function of end-to-end latency of query processing on a single node compared to periodic request (PR) and Flink. Our approach outperforms both Flink and PR, with lowest $\mean{x}(UCL)= 7.3 ms$ for the window query and highest $\mean{x}(UCL)= 14.48 ms$ for the predict query.}
\label{fig:evallat_1}
\end{figure*}

\begin{table*}[]
\scriptsize
\begingroup
\setlength{\tabcolsep}{4.3pt} 
\begin{tabular}{l|l|l|l|l|l|l|l|l|l|l|}
\cline{2-11}
                                            & \multicolumn{2}{l|}{\textbf{window}}           & \multicolumn{2}{l|}{\textbf{filter}}           & \multicolumn{2}{l|}{\textbf{join}}             & \multicolumn{2}{l|}{\textbf{heatmap}}          & \multicolumn{2}{l|}{\textbf{predict}}          \\ \hline
\multicolumn{1}{|l|}{\textbf{System}}       & \textit{mean}  & \textit{quantiles (90,95,99)} & \textit{mean}  & \textit{quantiles (90,95,99)} & \textit{mean}  & \textit{quantiles (90,95,99)} & \textit{mean}  & \textit{quantiles (90,95,99)} & \textit{mean}  & \textit{quantiles (90,95,99)} \\ \hline
\multicolumn{1}{|l|}{\textbf{Flink}}        & 2.97           & 0.15, 0.16, 0.18              & 0.329          & 0.155, 0.17, 0.191            & 10.627         & 20.64, 41.52, 41.53           & 2.098          & 4.09, 4.128, 4.167            & 12.44          & 0.34, 0.36, 0.39              \\ \hline
\multicolumn{1}{|l|}{\textbf{INetCEP\_PR}}  & 1.51           & 1.9, 1.96, 2.01               & 1.41           & 1.85, 1.92, 1.97              & 5.54           & 5.94, 5.99, 6.03              & 1.46           & 1.75, 1.85, 1.91              & 1.49           & 1.88, 1.95, 1.99              \\ \hline
\multicolumn{1}{|l|}{\textbf{INetCEP\_UCL}} & \textbf{0.007} & \textbf{0.009, 0.010, 0.014}  & \textbf{0.010} & \textbf{0.014, 0.017, 0.020}  & \textbf{0.011} & \textbf{0.017, 0.02, 0.025}   & \textbf{0.010} & \textbf{0.013, 0.015, 0.019}  & \textbf{0.014} & \textbf{0.019, 0.02, 0.024}   \\ \hline
\end{tabular}
\endgroup
\setlength{\abovecaptionskip}{0pt}
\setlength{\belowcaptionskip}{-15pt}
\caption{End-to-end latency in seconds for five queries. \system is $888 \times$ (best case: predict query) and $32 \times$ (worse case: filter query) better compared to the Flink CEP system as well as $503 \times$ (best case: join query) and $141 \times$ (worst case: filter query) better compared to the PR approach.}
\label{tab:lat_c}
\end{table*}

\subsubsection{Event Loss during Forwarding}
To further investigate the packet/event loss during forwarding, we measured the total loss rate for UCL, PR, and Flink. In \Cref{table:plr}, we present the \emph{mean}, \emph{min, max}, and \emph{quantiles (90, 95, 99)} values of the loss rate (in \%). For a low event rate, we observe a slight loss in events for the PR approach, which is expected due to the equal number of \Interest packets sent to retrieve the data stream. For a high event rate of 50,000 events per second, the PR approach has more event loss with a mean up to $61.18\%$, due to the UDP buffer overflow as well as the flood of \Interest and \Data packets.

Interestingly, we also observe event loss in Flink with Kafka as the data producer. Although this contradicts to Kafka being highly resilient, Kafka depends on many configuration parameters to prevent event loss, which are potentially conflicting and hardly optimizable\footnote{Data loss in Flink with Kafka. \url{https://ci.apache.org/projects/flink/flink-docs-stable/dev/connectors/kafka.html} [Accessed on 1.11.2020]}. With the default configuration parameters, we experienced that there is event loss even while using Flink and Kafka, as summarized in \Cref{table:plr}.

Nevertheless, our approach showed \emph{no event loss} at all, thanks to the flow control mechanism and the in-network caching mechanism of ICN.

\subsection{Performance Evaluation of \system Query Engine}
In this section, we evaluate the performance of the \system query engine as described in \Cref{subsec:QE}. We first evaluate on a single node setup  similar to the section above with the queries (cf. \Cref{tab:config-parameters}). Next, we provide a distributed evaluation of query engine, where operator placement is involved as well.

\subsubsection{Centralized Processing} \label{subsec:eq2}
We measure the performance of the five queries on a single node using the Google Cloud c2 instance. Here, the producer sends the data stream to the broker while the consumer registers the query by means of a \AddQInterest packet specifying the queries~\ref{query:window}--\ref{query:predict}. We re-implemented the same queries on the Flink engine and measured its performance as well.
Although in centralized processing, there is only one possible broker for the placement of the operator, the process of parsing, deployment and processing is time-consuming. In the PR approach, for each trigger to a query, a new \Interest packet is sent, and the placement is triggered each time a complex event is generated. For example, in the window query \ref{query:window}, when a new event arrives, a window of 4s is released.
This means, (i) the \Interest packet is sent every time\footnote{For the experiment, we have set the rate at which the consumer sends the \Interest equal to the producer sending rate, but in the real world, this is another problem with PR approach.} a new event arrives, (ii) the query is parsed, placed, and processed, and (iii) finally, the complex event is released. For this reason, we see that the performance of the PR approach is worse, which is a consequence of the pull-based approach.

\textbf{Impact on Latency.}
In \Cref{fig:evallat_1}, we show the CDF of end-to-end latency for the different queries. Similar to the forwarding results, the UCL approach is $30 \times$ better in latency than Flink. For example, the filter query UCL has a mean delay of $10.72 ms$, while Flink has $329.47 ms$. Most importantly, for the join query we see a substantial gain (about $100 \times$ better). We investigated the cause of high latency for query processing in Flink and we saw that the serialization and deserialization of data stream takes a significant amount of time. Furthermore, Flink suffers from the drawback of Kafka's pull-based approach, although with the batching mechanism Flink is better than PR in some queries. The PR approach performs better than Flink for some queries, for example, the heatmap query has a mean delay of $1.4 s$ in comparison to Flink, which has a mean delay of $2.0 s$. In essence, we outperform the PR approach by a substantial factor of $100 \times$.
In \Cref{tab:lat_c}, we summarize the mean delay and its quantiles (90, 95, 99).
As we have seen in the previous evaluations, the UCL approach exceeds the performance of Flink by a high factor, so we will proceed with only the PR approach for our comparison in the distributed evaluations.

\subsubsection{Distributed Processing} \label{subsec:eq3}
Similar to the single node evaluations, we measure the performance with the same five queries in the distributed evaluations as well. We use different topologies and numbers of nodes to investigate the behavior of operator placement and processing of different queries as well as to show the applicability of our approach to a wide range of scenarios. As shown in \Cref{fig:topo}b, we will first use the Manhattan graph
to measure the performance of our approach. In the topology, we consider a cloud server, edge switches, and end nodes to process the queries. We used the AWS instances for the distributed evaluation with seven nodes, as illustrated in \Cref{fig:topo}b on the left. 
All AWS instances run the NFN compute server, ccn-lite relay, and the nodes communicate using the CCN protocol.

\begin{figure*}[t!]
\centering
\includegraphics[width=0.75\linewidth]{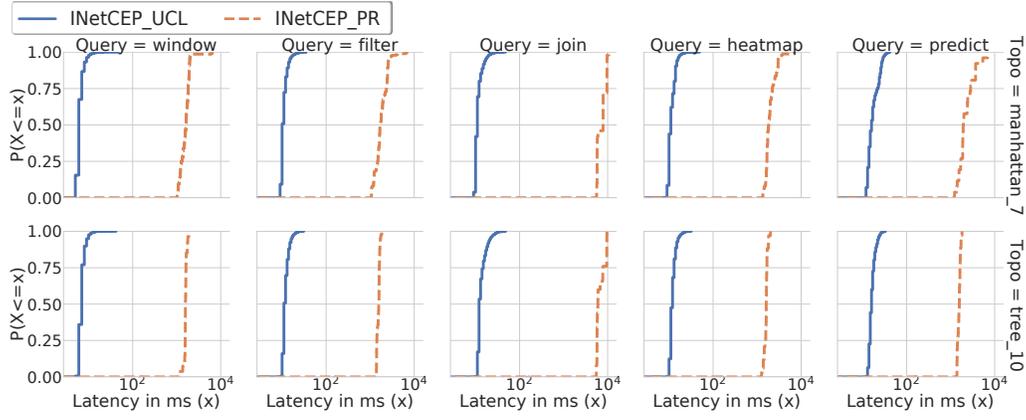}
\setlength{\abovecaptionskip}{5pt}
\setlength{\belowcaptionskip}{0pt}
\caption{Cumulative distribution function of end to end latency for query processing using an Manhattan graph (top) and tree topology (bottom) with 7 and 10 nodes, respectively. }
\label{fig:evallat_both}
\end{figure*}

\begin{table*}[]
\scriptsize
\begingroup
\setlength{\tabcolsep}{4pt} 
\begin{tabular}{@{}l|l|l|l|l|l|l|l|l|l|l|}
\cline{2-11}
                                                 & \multicolumn{2}{l|}{\textbf{window}}           & \multicolumn{2}{l|}{\textbf{filter}}           & \multicolumn{2}{l|}{\textbf{join}}             & \multicolumn{2}{l|}{\textbf{heatmap}}          & \multicolumn{2}{l|}{\textbf{predict}}          \\ \hline
\multicolumn{1}{|l|}{\textbf{System}}            & \textit{mean}  & \textit{quantiles (90,95,99)} & \textit{mean}  & \textit{quantiles (90,95,99)} & \textit{mean}  & \textit{quantiles (90,95,99)} & \textit{mean}  & \textit{quantiles (90,95,99)} & \textit{mean}  & \textit{quantiles (90,95,99)} \\ \hline
\multicolumn{1}{|l|}{\textbf{INetCEP\_PR\_M}}  & 1.634          & 1.951, 1.98, 6.31             & 1.883          & 2.559, 2.702, 5.354           & 7.424          & 9.646, 9.652, 10.678          & 2.061          & 2.985, 2.991, 6.926           & 2.503          & 3.702, 5.38, 6.939            \\ \hline
\multicolumn{1}{|l|}{\textbf{INetCEP\_UCL\_M}} & \textbf{0.006} & \textbf{0.008, 0.009, 0.012}  & \textbf{0.011} & \textbf{0.014, 0.015, 0.021}  & \textbf{0.011} & \textbf{0.015, 0.017, 0.023}  & \textbf{0.011} & \textbf{0.014, 0.016, 0.020}  & \textbf{0.018} & \textbf{0.027, 0.029, 0.035}  \\ \hline
\multicolumn{1}{|l|}{\textbf{INetCEP\_PR\_T}}  & 1.59           & 1.764, 1.789, 1.886           & 1.542          & 1.664, 1.698, 1.797           & 6.955          & 9.527, 9.56, 9.594            & 1.599          & 1.768, 1.82, 1.943            & 1.556          & 1.704, 1.725, 1.784           \\ \hline
\multicolumn{1}{|l|}{\textbf{INetCEP\_UCL\_T}} & \textbf{0.007} & \textbf{0.009, 0.010, 0.012}  & \textbf{0.012} & \textbf{0.015, 0.017, 0.021}  & \textbf{0.013} & \textbf{0.019, 0.022, 0.030}  & \textbf{0.012} & \textbf{0.015, 0.016, 0.021}  & \textbf{0.016} & \textbf{0.021, 0.023, 0.027}  \\ \hline
\end{tabular}
\endgroup
\setlength{\abovecaptionskip}{5pt}
\setlength{\belowcaptionskip}{-20pt}
\caption{End-to-end latency in seconds for distributed evaluations of Periodic Request (PR) and Unified Communication Layer (UCL). Table summarizes the \emph{mean} and \emph{quantiles (90,95,99)}. Here, INetCEP\_PR\_M stands for the PR approach with a Manhattan topology and INetCEP\_PR\_T stands for PR with a tree topology. Similar to the single node results, we supersede the PR approach by a factor of $674 \times$ (best case for join query) and $128 \times$ (worst case for filter query).}
\label{tab:lat_d}
\end{table*}

\textbf{Manhattan Graph.}
In the Manhattan graph, the end node $n6$ acts as a producer, $n5$ acts as a consumer that triggers the query, and nodes $n1, \dots, n4, n7$ acts as brokers, respectively. In \Cref{fig:evallat_both}, we present the CDF on the latency measurements for \system with UCL as a communication strategy and the query engine as well as \system with PR as a communication strategy and the query engine. In the PR approach, similar to the centralized evaluations, the query parser and placement is invoked each time the complex event is retrieved. In the top figure, we show the results for the Manhattan graph. Our approach clearly outperforms the naive pull strategy also in distributed query processing. It has the lowest mean delay $\mean{x} (UCL) = 6.46 ms$ for the window query and the highest $\mean{x} (UCL) = 18.08 ms$ for the predict query due to its high complexity~\cite{Martin2014}.

\textbf{Tree Topology.}
Next, we repeat the evaluations on the line and tree topology, as illustrated in Fig.~\ref{fig:topo}b (top and bottom on right) using CORE and MACI to set up the topology and to run evaluations in multiple nodes. We used AWS cloud instances as before and repeated the five queries. We observe identical results for the line and tree topology. Hence, we only present the results of the tree topology that shows only a slight difference, as shown in \Cref{fig:evallat_both} (bottom). Our approach is substantially better than the  PR approach with a lowest mean delay of $\mean{x} (UCL) = 7.07 ms$ for the window query and a highest $\mean{x} (UCL) = 16.65 ms$ for the predict query, consistent with the results before. In \Cref{tab:lat_d}, we summarize the mean and quantiles (90, 95, 99) for the delay observed for the two topologies. While the query processing results are similar, we expect the placement time to be different in the two topologies, as evaluated below.
\begin{figure}[t!]
\centering
\includegraphics[width=0.8\linewidth]{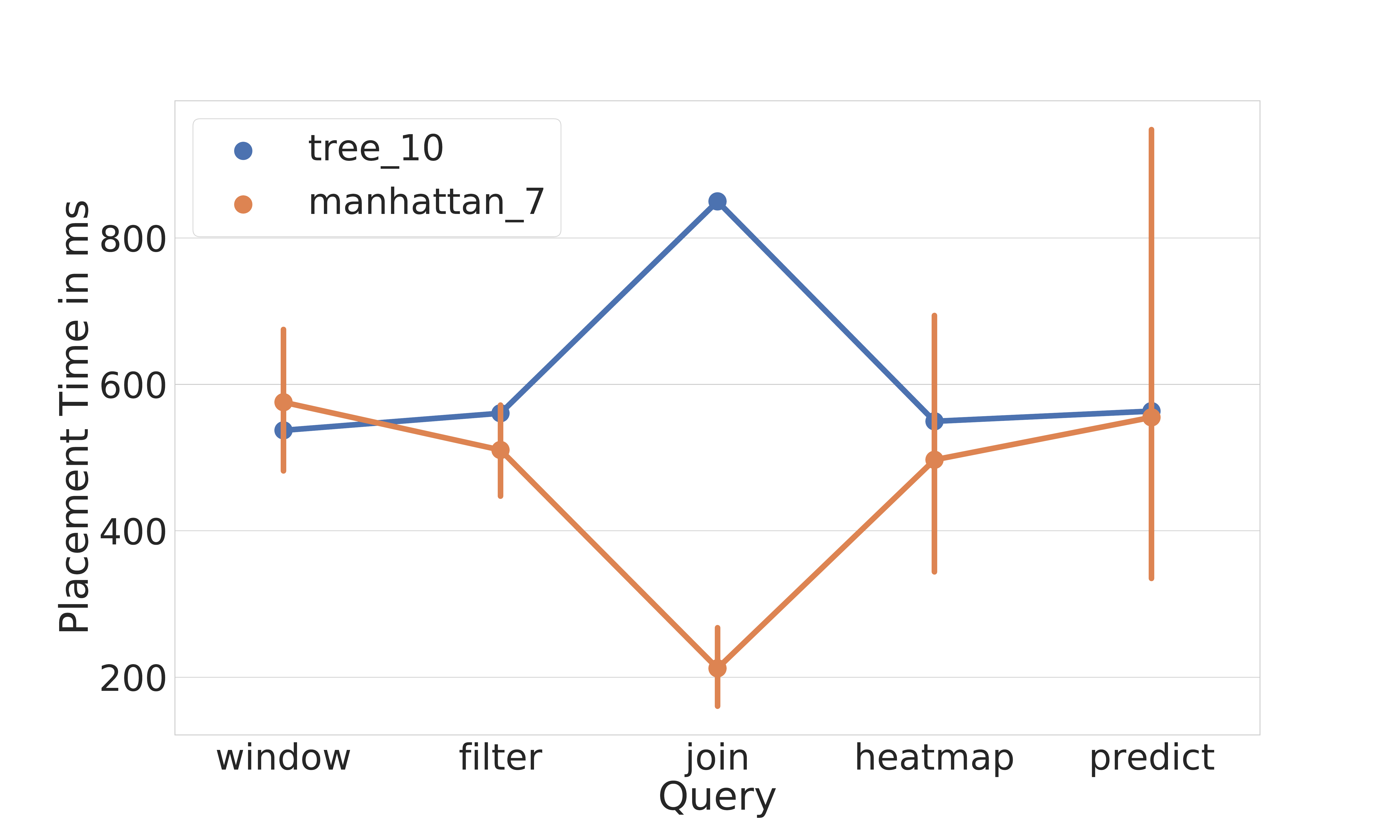}
\setlength{\abovecaptionskip}{0pt}
\setlength{\belowcaptionskip}{0pt}
\caption{A high number of operators takes more time for placement in a high number of nodes, but still in the order of a few milliseconds thanks to parallel processing.}
\label{fig:evallat_pltime}
\end{figure}

\begin{figure}[t!]
\centering
\includegraphics[width=\linewidth]{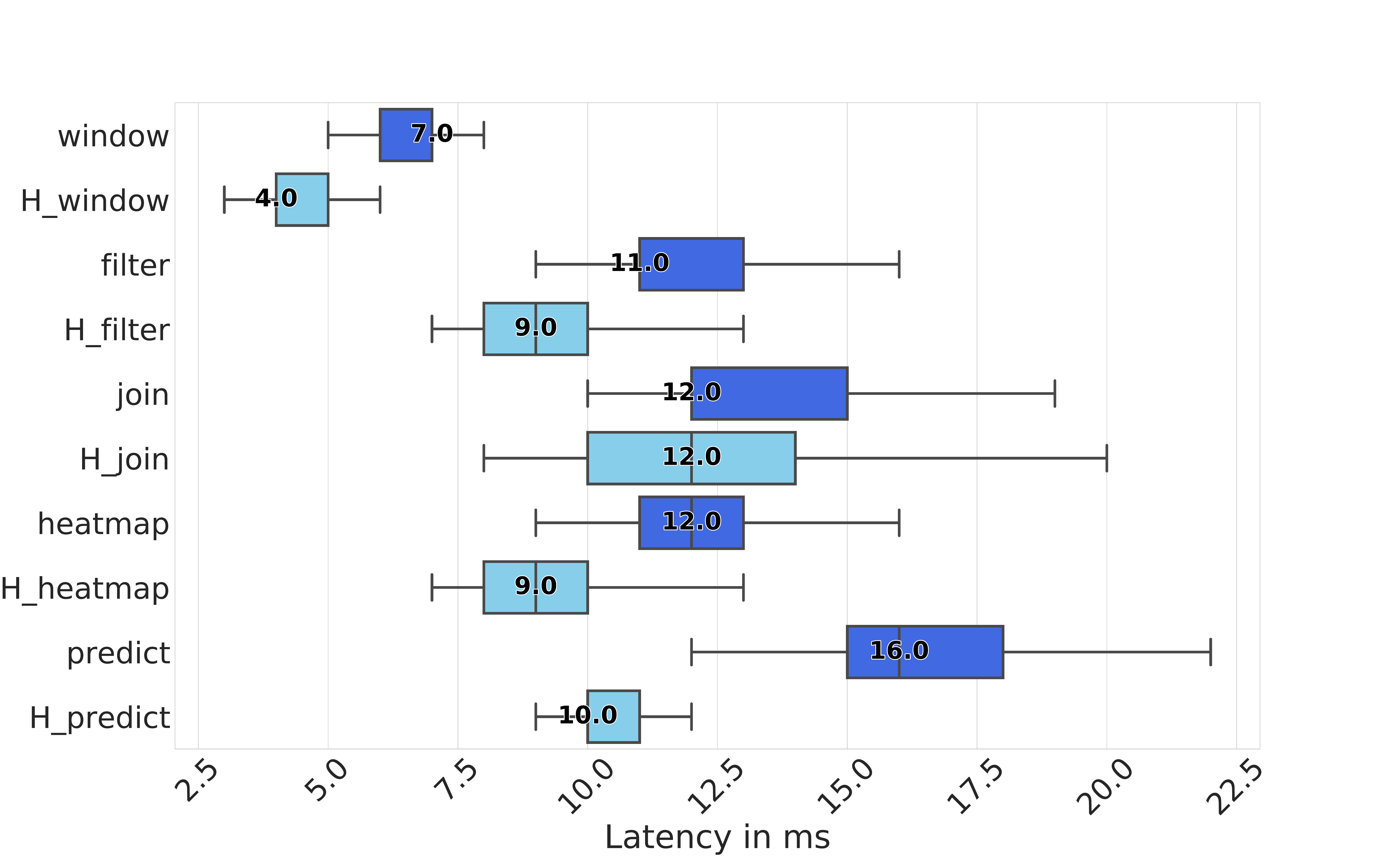}
\setlength{\abovecaptionskip}{-2pt}
\setlength{\belowcaptionskip}{0pt}
\caption{End-to-end latency of query processing in the tree topology with 10 nodes. Comparing application level queries with hybrid (denoted as "H\_Q", where $Q \in \{window, filter, join, heatmap, predict\}$) -- with a window operator deployed and processed in the network layer.}
\label{fig:evallat_hybrid}
\end{figure}
\textbf{Placement Time.}
We now analyze the placement time for the different topologies. It is defined as the time required to find an optimal path for the deployment of operators. We consider the Manhattan graph with 7 nodes and the tree topology with 10 nodes similar to the query processing evaluations. In \Cref{fig:evallat_pltime}, we present a point plot of the placement time with the median represented by the point and 95\% confidence interval represented by error bars for the five queries. As expected, due to high number of nodes and the paths to be discovered, the time to place operators in the tree topology is higher than in the Manhattan graph, but there are higher confidence intervals in the Manhattan graph in contrast to the tree topology. Further investigations indicate that when farther away nodes are used for placement, it took longer to place than with nearer nodes. Furthermore, the tree topology was evaluated using the CORE emulator where everything ran on a single node with virtualization for the 10 nodes using linux namespaces. Hence, due to no real network latencies, there was no or a minimum difference in the placement time, as indicated by the negligible confidence interval for the tree topology.

Another interesting observation is the time taken to place join query. In the tree topology, it takes almost triple the time in comparison to the Manhattan graph. This is because a higher number of operators is involved in a join query (6 in total) than in other queries. The time to discover paths in the tree topology is also higher than in the Manhattan graph and hence we see a larger difference in placement time with join queries than for other queries.

\textbf{Hybrid Processing.}
To further benefit from the performance of the data plane of CCN, we implemented a window operator in the CCN-lite implementation of a CCN router (network layer). While all the other operators executes on a NFN compute server, we investigated the performance of the five queries when the window operator executes on a CCN router. In \Cref{fig:evallat_hybrid}, we present the results in a box plot for the end-to-end latency of queries running completely in the application layer and partially running in the network layer (or hybrid and denoted as H\_Q). A box plot represents a five number summary of the latency measurements, minimum, first quartile (Q1), median, third quartile (Q3), and maximum. The vertical line in between sometimes appearing close to Q1 is the median, we also report the median values written as an overlay in the figure. The left bound represents the minimum, the start of the box plot Q1, the end represents Q3 and the right bound represents the maximum. We have excluded the outliers from the boxplot.

Hence, we clearly observe the benefit of moving the query operation from the application layer to the network layer, as seen in the plot the hybrid queries always outperforms the queries running at the application level. This is due to the round trip to the NFN compute server. The minimum median is for the hybrid window query $4 ms$, which completely runs on top of a CCN router.

\subsubsection{Impact on Accuracy}
In this paragraph, we evaluate the impact of event loss on the accuracy in the delivery of the complex events. As indicated in the smart plug example for power grid providers in \Cref{sec:problem}, such industrial IoT applications need to be highly robust and hence cannot tolerate failures.  Event loss can result in false negatives and false positives in complex events, which  eventually can cause heavy loss if not complete failure. For this reason, we measure the accuracy in the delivery of complex events for the five queries. We measure it in terms of the F1-score that better represents accuracy in the results. In \Cref{tab:accuracy}, we present the F1-score of the two queries heatmap and predict that are more prone to errors and are implicitly complex. As a consequence of event loss, the PR approach either misses or leads to many false positives and negatives, hence we see a drop in accuracy. Similar observation is seen for Flink since it suffers from data loss as well. In contrast, \system does not suffer from any event loss and hence correctly identifies all the complex events with a F1-score of 100\%, thanks to the lossless flow control mechanism in combination with the placement strategy.

\begin{table}[]
\scriptsize
\begin{tabular}{l|l|l|l|l|}
\cline{2-5}
                                                                                                   & \multicolumn{2}{l|}{\textbf{heatmap}}         & \multicolumn{2}{l|}{\textbf{predict}}         \\ \hline
\multicolumn{1}{|l|}{\textbf{System}}                                                              & \textit{mean} & \textit{quantiles (90,95,99)} & \textit{mean} & \textit{quantiles (90,95,99)} \\ \hline
\multicolumn{1}{|l|}{\textbf{\begin{tabular}[c]{@{}l@{}}Centralized, \\ INetCEP\_PR\end{tabular}}} & 54.36        & 69.80, 70.90, 71.78        & 48.70        & 46.62, 51.88, 56.09        \\ \hline
\multicolumn{1}{|l|}{\textbf{\begin{tabular}[c]{@{}l@{}}Distributed, \\ INetCEP\_PR\end{tabular}}} & 66.66        & 66.66, 66.66, 66.66        & 66.66        & 66.66, 66.66, 66.66        \\ \hline
\end{tabular}

\caption{Accuracy in complex events generated by the heatmap and predict queries for the periodic request approach.  \system does not suffer from any event loss and hence correctly identifies all complex events with a F1-score of 100\%.}
\label{tab:accuracy}
\end{table}

%% file: sections/discussion.tex
\section{Conclusion}\label{sec:conclusion}
We presented a novel in-network processing architecture, \system, that implements a unified communication layer for co-existing consumer-initiated (pull-based) and producer-initiated (push-based) interaction patterns. 
In this way, a wide range of \ac{IoT} applications with push notifications and request-reply interactions are supported.
With the proposed meta query language, we can express interest in aggregated data that is resolved and processed in a distributed manner in an IoT network.
In our evaluation, we showed that \system is more than $15 \times$ faster in terms of forwarding events than the state-of-the-art CEP system Flink.
Furthermore, the delivery and processing of complex queries on the mentioned datasets is about \textbf{$32\times$} faster than \emph{Flink} and more than $100\times$ faster than a naive \emph{ pull-based} reference approach while maintaining $100\%$ accuracy.
There are several areas for future work. 
For example, extending \system to run on the CCN-lite kernel module promises high performance benefits for our unified communication layer\footnote{In our early efforts, we have updated CCN-lite to support the latest Linux kernel. \url{https://github.com/cn-uofbasel/ccn-lite/pull/375}. [Accessed on 1.11.2020]} when using the Linux kernel, similar to other work on CEP in the kernel~\cite{Graubner18}.
Another promising area of future work is to generate an optimal operator graph, \eg based on operator \emph{selectivity} or by partitioning the operator graph by performing query optimization~\cite{Cao2013}. The placement module should be extended to support other decentralized solutions and/or other QoS metrics~\cite{Cardellini2016} like throughput and availability. Finally, eliminating DDoS attacks for push-based communication is another research direction.

\section*{Acknowledgements}
This work is funded by the German Research Foundation (DFG) as part of the project C2, A3 and C5 within the Collaborative Research Center (CRC) SFB 1053 -- MAKI.